\documentclass[10pt,conference]{IEEEtran}
\usepackage{cite}
\usepackage{amsmath,amssymb,amsfonts}
\usepackage{algorithmic}
\usepackage{graphicx}
\usepackage{textcomp}
\usepackage{xcolor}
\usepackage[hyphens]{url}
\usepackage{fancyhdr}
\usepackage{hyperref}

\usepackage{booktabs}
\usepackage{multirow}
\usepackage{makecell}
\usepackage[inkscapelatex=false]{svg}
\usepackage{subcaption}

\usepackage{caption}
\usepackage{pifont}
\usepackage[dvipsnames]{xcolor} 
\usepackage{enumitem}

\title{Modeling and Optimizing Performance Bottlenecks for Neuromorphic Accelerators}

\author{
\IEEEauthorblockN{
Jason Yik\IEEEauthorrefmark{1}, 
Walter Gallego Gomez\IEEEauthorrefmark{2},
Andrew Cheng\IEEEauthorrefmark{1}, 
Benedetto Leto\IEEEauthorrefmark{2}, 
Alessandro Pierro\IEEEauthorrefmark{3}\IEEEauthorrefmark{4},
Noah Pacik-Nelson\IEEEauthorrefmark{5}\IEEEauthorrefmark{6}, 
\\
Korneel Van den Berghe\IEEEauthorrefmark{7}, 
Vittorio Fra\IEEEauthorrefmark{2}, 
Andreea Danielescu\IEEEauthorrefmark{5}\IEEEauthorrefmark{8}, 
Gianvito Urgese\IEEEauthorrefmark{2}, 
Vijay Janapa Reddi\IEEEauthorrefmark{1}
}\\
\IEEEauthorblockA{
\IEEEauthorrefmark{1}Harvard University
\IEEEauthorrefmark{2}Politecnico di Torino
\IEEEauthorrefmark{3}Intel
\IEEEauthorrefmark{4}LMU Munich
\IEEEauthorrefmark{5}Accenture Labs
\IEEEauthorrefmark{6}BootLoop AI
\IEEEauthorrefmark{7}TU Delft
\IEEEauthorrefmark{8}Wordly
}
}

\begin{document}
\pagestyle{plain}      
\maketitle
\thispagestyle{plain}  
\vspace{-1cm}


\begin{abstract}
Neuromorphic accelerators offer promising platforms for machine learning (ML) inference by leveraging event-driven, spatially-expanded architectures that naturally exploit unstructured sparsity through co-located memory and compute. 
However, their unique architectural characteristics
create performance dynamics that differ fundamentally from conventional accelerators. 
Existing workload optimization approaches for neuromorphic accelerators rely on aggregate network-wide sparsity and operation counting, but the extent to which these metrics actually improve deployed performance remains unknown.
This paper presents the first comprehensive performance bound and bottleneck analysis of neuromorphic accelerators, revealing the shortcomings of the conventional metrics and offering an understanding of what facets matter for workload performance.
We present both theoretical analytical modeling and extensive empirical characterization of three real neuromorphic accelerators: Brainchip AKD1000, Synsense Speck, and Intel Loihi 2. 
From these, we establish three distinct accelerator bottleneck states, memory-bound, compute-bound, and traffic-bound, and identify which workload configuration features are likely to exhibit these bottleneck states.
We synthesize all of our insights into the floorline performance model, a visual model that identifies performance bounds and informs how to optimize a given workload, based on its position on the model.
Finally, we present an optimization methodology that combines sparsity-aware training with floorline-informed partitioning.
Our methodology achieves substantial performance improvements at iso-accuracy: up to 3.86$\times$ runtime improvement and 3.38$\times$ energy reduction compared to prior manually-tuned configurations. 
This work lays the groundwork for system-level optimization of neuromorphic accelerators, and it provides architectural insights and a systematic optimization framework that enables principled performance engineering for this class of architectures.
\end{abstract}

\section{Introduction}
Neuromorphic accelerators employ a distinct architectural approach for executing sparse neural network inference, characterized by spatially-expanded designs where each logical neuron maps to a dedicated physical compute unit on-chip. This contrasts with conventional accelerators that time-multiplex logical neurons across shared arithmetic units~\cite{khacef18confronting, schuman2017surveyneuromorphiccomputingneural, roy2019towards, basu22snnhwreview}. The neuromorphic approach co-locates memory with processing units (neurocores) and leverages event-driven execution to exploit activation sparsity without requiring structured sparsity patterns. Recent deployments have demonstrated substantial performance and energy benefits for specific sparse workloads compared to edge GPU baselines~\cite{shrestha2024efficient, meyer2024diagonalstructuredstatespace, pierro2025acceleratinglinearrecurrentneural,abreu2025neuromorphicprinciplesefficientlarge, yik2025neurobench}.

However, the unique architectural characteristics of neuromorphic accelerators create performance dynamics that are poorly understood. While recent algorithmic work has focused on optimizing neural networks for neuromorphic deployment~\cite{shi2024towards, chakraborty2024sparse, shen23esl-snn,pierro2025acceleratinglinearrecurrentneural, yan2022sparsityregularization, na2022autosnnenergyefficientspikingneural, imec2024optimizingseneca, nazeer24languagesp2}, these efforts rely on high-level performance proxies, primarily network-wide sparsity and aggregate operation counting, without testing how well such proxies actually inform deployed performance.
This gap is critical: no systematic analysis has established which factors actually drive neuromorphic accelerator performance and how to optimize for them.

In this paper, we address this gap by presenting a ``bound and bottleneck'' analysis~\cite{lazowska1984quantitative} of neuromorphic accelerator performance, mirroring methodology which produced the widely-used roofline performance model for conventional architectures~\cite{williams09roofline}. Our analysis aims to answer three key questions about neuromorphic systems: (1) What core operations bound and bottleneck performance? (2) Which workload configurations result in different bounds and bottlenecks? (3) How does one optimize a given workload using an understanding of its bottleneck and performance bounds?

Our study proceeds by first using a simple analytical architecture and workload model to provide theoretical insights into how memory, compute, and traffic operations scale with workload sparsity and parallelization on a neuromorphic chip. Next, we verify and substantiate the insights by extensively profiling three real neuromorphic accelerators: the edge-focused Brainchip AKD1000~\cite{brainchipakd1000}, the event-camera specialized Synsense Speck~\cite{man24speck}, and the research-oriented Intel Loihi~2~\cite{loihi2techbrief}. Based on the modeling and profiling insights, we synthesize the floorline model, an analog to the roofline model, which visually indicates the performance bounds of a neural network architecture and informs how to optimize any trained network instantiation.


Our analytical modeling and profiling reveal several important insights about neuromorphic accelerator performance. Most critically, we find that conventional network-wide performance proxies are insufficient for neuromorphic architectures due to neurocore-level load imbalance; instead, neurocore-aware metrics are necessary for understanding whether performance will improve. We also observe that current neuromorphic implementations show limited ability to exploit weight sparsity for convolutional networks (CNNs), suggesting that recently proposed CNN weight pruning approaches~\cite{shi2024towards, shen23esl-snn, chakraborty2024sparse} may require architectural modifications to be effective. For bounds and bottlenecks, we identify memory accesses during synaptic operations (synops) as the usual dominant workload cost, consistent with prior circuit-level analysis~\cite{Furber_2016, dampfhoffer23aresnn, tang2023openboxdigitalneuromorphic}. However, we importantly uncover that certain workloads can be bound by neuron computes or message traffic, necessitating different optimization approaches than the synops-bound state.

Based on these observations, we develop a two-stage optimization methodology that applies our performance model insights. The first stage co-optimizes network accuracy and sparsity during training to leverage the fundamental sparsity benefits in neuromorphic architectures. The second stage uses our architecture-aware floorline performance model to iteratively optimize neurocore partitioning and mapping. Applied to our three platforms at iso-accuracy operating points, this approach achieves notable performance improvements over baseline configurations: up to 4.29$\times$ runtime and 4.36$\times$ energy improvements from sparsity optimization, with additional gains of up to 1.83$\times$ runtime and 1.52$\times$ energy from architecture-aware optimization on Loihi~2. The combined two-stage optimization yields up to 3.86$\times$ runtime and 3.38$\times$ energy improvements compared to prior manually-tuned configurations for the studied workloads.


In summary, this work studies the performance bounds and bottlenecks of neuromorphic accelerators with the following contributions:
\begin{itemize}
    \item We present analytical modeling to provide insights for how network sparsity and parallelization configurations affect memory, compute, and traffic bottlenecks.
    \item We empirically demonstrate that real neuromorphic accelerator workloads can be in any of the three bottleneck states, under configurations expected by the analytical model, and characterize the performance boundaries that result from the bottlenecks.
    \item We propose the floorline performance model, a visual model which synthesizes the relationships between performance bounds and bottlenecks, and informs how to understand and optimize a workload's performance given its location on the model.
    \item We develop and validate a generic two-stage optimization methodology that utilizes sparsity-aware training and floorline-informed partitioning, achieving up to 3.86$\times$ runtime and 3.38$\times$ energy improvements over prior configurations.
\end{itemize}

\section{Background}
\label{sec:background}

This section provides the architectural background needed to understand neuromorphic accelerator performance characteristics and the unique challenges they present for performance modeling. In addition, we review related work and identify the research gaps that our paper uniquely addresses.

\subsection{Neuromorphic Accelerator Architecture}

\begin{figure}[h]
\centering
    \includegraphics[width=0.47\textwidth]{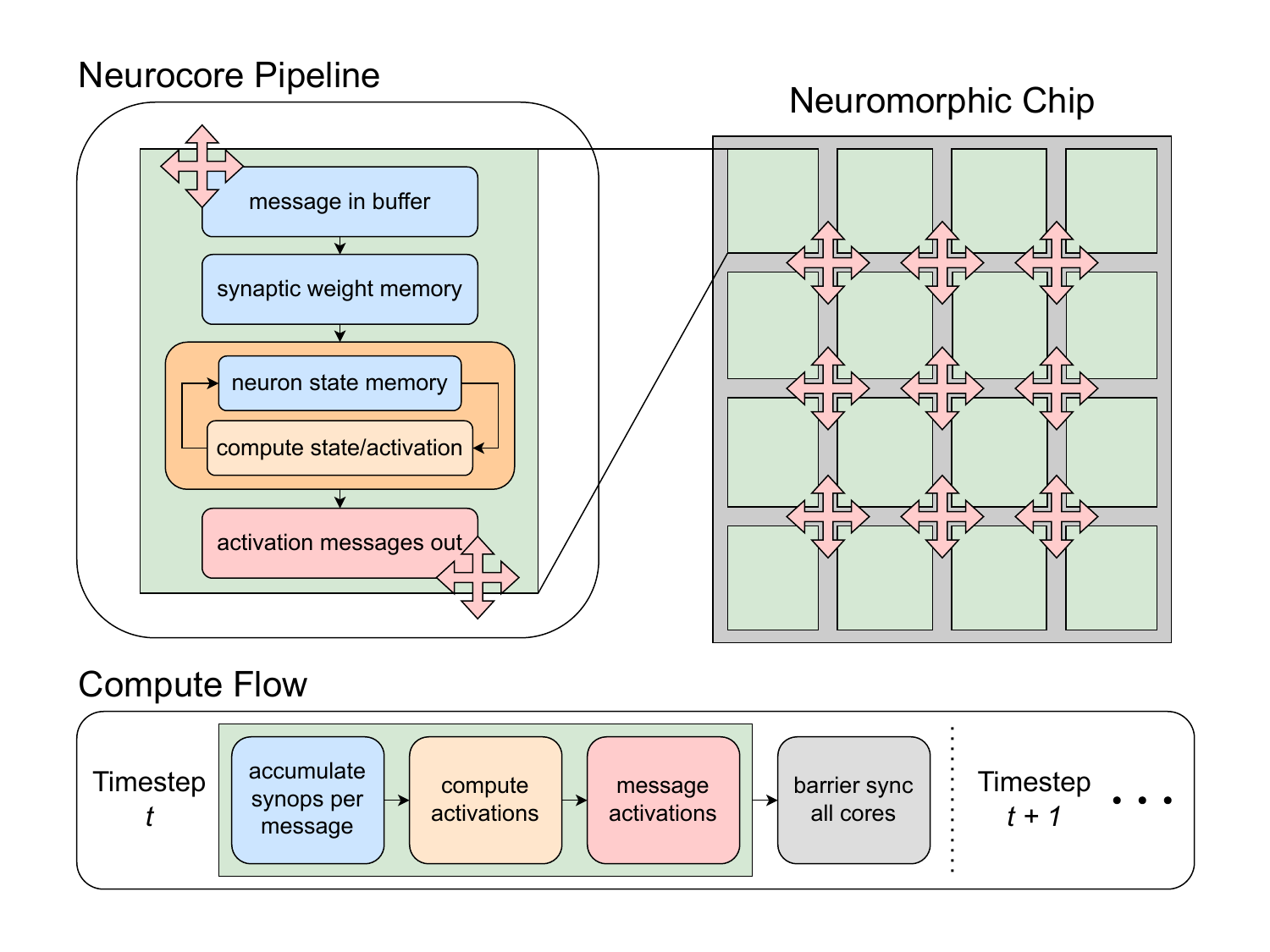}
    \caption{The general macro-architecture and compute flow of neural network execution on neuromorphic accelerators. Blue components denote co-located memory, orange denotes compute, and red denotes inter-core communication via the NoC. 
    }
    \label{fig:general_architecture}
\end{figure}

Figure~\ref{fig:general_architecture} illustrates the neuromorphic accelerator architecture, which is applicable across current designs~\cite{davies18loihi, loihi2techbrief, brainchipakd1000, man24speck, akopyan15truenorth, furber14spinnaker, Mayr2019spinnaker2, tang2023seneca, pei2019tianjic, moreira20neuronflow}\footnote{Since the term `neuromorphic' can be broadly applied, many other computing systems have been proposed and labeled under the term, such as compute-in-memory and crossbar array architectures~\cite{schuman2017surveyneuromorphiccomputingneural, roy2019towards}. This study focuses on the macro-architecture shown in Figure~\ref{fig:general_architecture}.}. The architecture operates at two levels: the chip-level organization (right) is an array of \textbf{neurocores} connected via a network-on-chip (NoC), and each neurocore (left) uses a pipeline structure to process neural computations.

The execution pipeline of each \textbf{neurocore} contains three memory components, a compute stage, and message output stage. The \textbf{message in buffer} stores input activation messages from other neurocores, the \textbf{synaptic weight memory} holds the weights for neurons assigned to this neurocore, and the \textbf{neuron state memory} maintains persistent state for stateful neuron models. 
The compute stage generates new activations based on the inputs, which are then formatted as sparse \textbf{activation messages} and sent via the NoC to downstream neurocores for processing in the subsequent timestep.

The bottom of Figure~\ref{fig:general_architecture} shows the temporal execution model where all neurocores operate in lockstep during discrete timesteps. Within each timestep, neurocores simultaneously: (1) accumulate the \textbf{synaptic operations (synops)} from each input message, (2) \textbf{compute activations}, (3) generate outgoing \textbf{activation messages}, and (4) participate in \textbf{barrier synchronization} before advancing to the next timestep. Synops for each message involves fetching weights, multiplying them by the input, and accumulating into a pre-activation for each neuron.

To map a neural network to the neuromorphic chip for inference, groups of neurons are logically \textbf{partitioned}, then physically \textbf{mapped} to a neurocore on the chip. For feedforward ML networks, all neurons in a neurocore belong to the same algorithmic network layer.

The \textbf{timestep duration} is the time required for the slowest neurocore to complete its assigned computation, establishing the accelerator's throughput bottleneck. Since neurocores operate in parallel but barrier synchronize at each timestep, the maximum processing time across all active neurocores determines overall accelerator performance. This creates a load balancing challenge where uneven computational workload distribution across neurocores directly impacts accelerator throughput, regardless of total computational capacity.

The event-driven execution paradigm naturally exploits sparsity through two mechanisms. \textbf{Weight sparsity} reduces memory access and synop accumulation requirements within each neurocore, while \textbf{activation sparsity} reduces NoC message traffic and downstream synop requirements. Unlike conventional accelerators that may require structured sparsity, neuromorphic accelerators can natively leverage unstructured sparsity through their per-message weight fetching and activation message passing approaches, though our analysis reveals architecture-specific limitations for certain sparsity types.

The neuromorphic spatially expanded approach, where each logical neuron maps to a dedicated physical location in a neurocore~\cite{khacef18confronting}, contrasts with conventional accelerators (GPUs, TPUs) that achieve parallelism through time-multiplexed execution across shared arithmetic units. Most critically for performance modeling, such exclusive use of on-chip co-located memory fundamentally alters the performance landscape compared to conventional accelerators. Unlike systems where off-chip memory bandwidth typically dominates performance, neuromorphic accelerators operate in a regime where on-chip memory access, local compute, and NoC communication costs are within the same order of magnitude~\cite{tang2023openboxdigitalneuromorphic, davies18loihi}. This creates a multi-dimensional performance space where the bottleneck can shift between intra-core synop memory access, intra-core activation computation, and inter-core communication traffic, depending on workload characteristics and system utilization.

\subsection{Related Work}

Prior work in neuromorphic performance falls into two main categories: performance estimation and hardware-software optimization. Performance estimation efforts have primarily relied on simulation-based approaches, including architectural simulators for specific designs~\cite{boyle2023sanafe, xie23spikenc} and micro-operation cost estimation through circuit-level simulation~\cite{tang2023openboxdigitalneuromorphic, Furber_2016, Timcheck_2023} or system micro-benchmarks~\cite{Ostrau2022, gomez2023first}. However, the architectural simulator approaches either focus on single systems or rely on simplified assumptions (e.g., fixed-duration timesteps), while the micro-operation cost estimates fail to capture system-level performance characteristics of parallel workload execution across neurocores.

Optimization efforts have explored partitioning and mapping strategies for neuromorphic platforms~\cite{urgese2016optimizing, lin18loihimapping, severa25benchmarkingpartitioning, balaji20spinemap, li20sneap, wu20rlmapping, song22dfsynthesizer, jin23mapping, yang24tianjiccompiling}, though most work remains limited to simulators without validation on real systems. Other recent works have demonstrated performance improvements through network sparsification~\cite{shrestha2024efficient, man24speck, nazeer24languagesp2, pierro2025acceleratinglinearrecurrentneural}, but lack systematic characterization of sparsity effects or cross-platform generalizability.

\begin{table}[t!]
    \centering
    \caption{Comparison of neuromorphic performance analysis features across prior work.}
    \label{tab:related_work_comparison}
    \resizebox{\columnwidth}{!}{%
    \begin{tabular}{@{}lcccccc@{}}
    \toprule
    & \textbf{\begin{tabular}[c]{@{}c@{}}Cross-\\Platform\end{tabular}} & \textbf{\begin{tabular}[c]{@{}c@{}}Real\\Hardware\end{tabular}} & \textbf{\begin{tabular}[c]{@{}c@{}}Arch-Aware\\Metrics\end{tabular}} & \textbf{\begin{tabular}[c]{@{}c@{}}System\\Bottlenecks\end{tabular}} & \textbf{\begin{tabular}[c]{@{}c@{}}Performance\\Framework\end{tabular}} & \textbf{\begin{tabular}[c]{@{}c@{}}Optimization\\Method\end{tabular}} \\ \midrule
    
    Boyle et al.~\cite{boyle2023sanafe} & \textcolor{green!60!black}{\ding{51}} & \textcolor{red}{\ding{55}} & \textcolor{red}{\ding{55}} & \textcolor{red}{\ding{55}} & \textcolor{red}{\ding{55}} & \textcolor{red}{\ding{55}} \\
    Tang et al.~\cite{tang2023openboxdigitalneuromorphic} & \textcolor{red}{\ding{55}} & \textcolor{red}{\ding{55}} & \textcolor{red}{\ding{55}} & \textcolor{red}{\ding{55}} & \textcolor{red}{\ding{55}} & \textcolor{red}{\ding{55}} \\
    Ostrau et al.~\cite{Ostrau2022}& \textcolor{green!60!black}{\ding{51}}  & \textcolor{green!60!black}{\ding{51}} & \textcolor{red}{\ding{55}} & \textcolor{red}{\ding{55}} & \textcolor{red}{\ding{55}} & \textcolor{red}{\ding{55}} \\
    Yang et al.~\cite{yang24tianjiccompiling} & \textcolor{red}{\ding{55}} & \textcolor{green!60!black}{\ding{51}} & \textcolor{green!60!black}{\ding{51}} & \textcolor{red}{\ding{55}} & \textcolor{red}{\ding{55}} & \textcolor{green!60!black}{\ding{51}} \\
    Pierro et al.~\cite{pierro2025acceleratinglinearrecurrentneural} & \textcolor{red}{\ding{55}} & \textcolor{green!60!black}{\ding{51}} & \textcolor{red}{\ding{55}} & \textcolor{red}{\ding{55}} & \textcolor{red}{\ding{55}} & \textcolor{green!60!black}{\ding{51}} \\
    \midrule
    \textbf{This Work} & \textcolor{green!60!black}{\ding{51}} & \textcolor{green!60!black}{\ding{51}} & \textcolor{green!60!black}{\ding{51}} & \textcolor{green!60!black}{\ding{51}} & \textcolor{green!60!black}{\ding{51}} & \textcolor{green!60!black}{\ding{51}} \\
    \bottomrule
    \end{tabular}%
    }
\end{table}
    
As shown in Table~\ref{tab:related_work_comparison}, existing work lacks cross-platform analysis using real hardware to characterize fundamental performance principles. 
Importantly, prior work has not investigated whether widely-used performance proxy metrics (network-wide sparsity, aggregate operation counts) truly reflect improvements in neuromorphic system performance, nor established a unifying performance framework capturing system bounds and bottlenecks. This paper addresses these gaps through the first systematic performance investigation with multiple real neuromorphic platforms.
\section{Analytical Modeling}
\label{sec:modeling}
In this section, we use a simple analytical model to theoretically inform the relationships between neuromorphic workload configuration settings and fundamental accelerator bottlenecks. We use this analytical modeling to guide our hardware profiling in Section~\ref{sec:characterization}.

The pipelined neurocore computing units have three core operations: (a) \textbf{synop accumulation}, (b) \textbf{activation computation}, and (c) \textbf{NoC activation messaging}. At the same time, the following features affect how many of each core operation is required: \textbf{weight sparsity} impacts the number of synop weight fetches and accumulations, \textbf{activation sparsity} impacts the number of synop inputs and NoC traffic, and \textbf{partitioning} affects the number of neurons in a neurocore that synops and activation computes are applied to, as well as the total neurocore utilization on-chip. Finally, neurocore \textbf{mapping} affects the traffic congestion and routing of activation messaging, though it does not affect the volume of total message traffic.

By varying workload configurations, the core operation counts will grow or shrink, and thus each of the three operations may bottleneck overall performance. Since the costs for synops, activation computes, and message routing are within an order of magnitude, we use asymptotic operation counts in order to compare the three~\cite{tang2023openboxdigitalneuromorphic, davies18loihi}.


\subsection{Model Formulation}

Our model examines one layer, $l_i$, in a fully connected network which has $N$ neurons. For simplicity, the previous and next layers connected to $l$, $l_{i-1}$ and $l_{i+1}$, also have $N$ neurons. 
The weight density in the synaptic connections is denoted $w$, such that the weight sparsity is $1-w$. We assume that $m$ is the message density of both $l_{i-1}$ and $l_i$, where the former impacts the input of $l_i$ and the latter impacts the input of $l_{i+1}$.



\subsection{Single Neurocore Per-Layer Operation}

We first uncover the `base case' bottleneck, by analyzing each layer mapped onto one neurocore without parallelization.

Accounting for activation sparsity, the number of inputs to $l_i$ is $mN$, and each input triggers the sparse read of $wN$ weights, leading to $mwN^2$ synops. Next, to count the minimum number of activation computes, we make the idealized assumption that an activation compute is only necessary if the neuron in $l_i$ received at least one synop. Under uniform sparsity, for each of the $mN$ inputs $a$, and for each of the $N$ neurons $b$:
\begin{align}
    p(a \text{ does not message } b) &= 1 - w \\
    \label{eq:receives_no_message}
    p(b \text{ receives no message)} &= (1-w)^{mN} \\ 
    p(b \text{ is messaged)} &= 1 - (1-w)^{mN}
\end{align}

Therefore, the minimum neurocore activation computes is $N \cdot (1-(1-w)^{mN})$. In practice, without extremely high sparsity, for the usual neuron dimensions of ML network layers, each neuron is expected to receive at least one message, so neuron updates and writeback remain $\sim O(N)$. Thus, the sparsity-aware operation costs are:
\begin{enumerate}[label=(\alph*)]
    \item Synops: $O(mwN^2)$
    \item Act. computes: $\sim O(N)$
    \item Messages traffic to $l_{i+1}$: $O(mN)$
\end{enumerate}

This indicates an operation bottleneck: \textbf{when sparsities are low, the quadratic $N^2$ scaling of synops will dominate.} As sparsity increases, and the synops are sufficiently scaled down, the system bottleneck may shift to activation computes or message traffic, which both scale linearly with $N$.

\subsection{Parallel Utilization of Multiple Neurocores}
\label{sec:voluntary_utilization}

Next, we introduce parallelization to the model, addressing the synops bottleneck by sharing the load among multiple neurocores. Here, we examine the case where a network layer is `voluntarily' partitioned to multiple neurocores. This type of utilization contrasts with the `involuntary' case, where greater utilization is forced as a layer's weight parameters scale beyond the memory limits of a neurocore, addressed next.

We retain neuron dimensions $N$ and define the number of neurocores assigned to each layer $C_{i-1}$, $C_i$, and $C_{i+1}$ of layers $l_{i-1}$, $l_i$, and $l_{i+1}$, respectively. 

For the $mN$ messages from $l_{i-1}$, by default, all must be broadcast such that each of the neurocores in $l_i$ will receive them to update their mapped neurons. Using an idealized hardware assumption that a message does not need to be sent to a neurocore if all of the corresponding synapses to the neurons are pruned, then the probability that an activation misses the $N/C_i$ neurons in a neurocore of $l_i$ is $(1-w)^{\frac{N}{C_i}}$. Again, in practice, this probability is very low, and the number of activations at each neurocore remains $\sim O(mN)$. Since (approximately) every message is duplicated to each neurocore in $l_{i+1}$, the partitioned operation costs are:
\begin{enumerate}[label=(\alph*)]
    \item Synops per neurocore: $O(mwN^2 / C_i)$
    \item Act. computes per neurocore: $\sim O(N / C_i)$
    \item Message traffic to $l_{i+1}$: $O(mNC_{i+1})$
\end{enumerate}

We identify two important implications of increasing neurocore utilization. First, \textbf{synops are processed in parallel across the active neurocores, meaning that this operation must be accounted for at the granularity of a neurocore, rather than in total.} Second, \textbf{increasing partitioning causes linearly increasing message traffic, and linearly decreasing neurocore synops.} This indicates that while partitioning can effectively address a synops bottleneck, it may shift the system to a traffic bottleneck.


\subsection{Forced Utilization from Network Width}
\label{sec:involuntary_utilization}

Finally, we examine the workload configuration case where a network layer's width has scaled up, which forces `involuntary' neurocore utilization due to the layer exceeding the synaptic memory capacity of one neurocore. Note that the prior `voluntary' parallelization can be applied on top of this forced utilization.

We re-define $N$ to be the maximum number of neurons that fit in a single neurocore. The new neuron dimension for all $l$ is $xN$, where $x > 1$. Due to the fully connected weight layers, scaling neuron dimension by $x$ leads to a $x^2$ scaling in number of synaptic weights, which leads to quadratic scaling in the number of neurocores necessary to minimally map each layer $l$: $C_{i-1, i, i+1} = O(x^2)$. 
This quadratic scaling of neurocores leads to a decrease in the number of neurons per neurocore in each layer, each neurocore maps $xN/x^2 = N/x$ neurons.

Counting operations for $l_i$, in each neurocore, each of the $mxN$ inputs fetches $wN/x$ synaptic weights for accumulation, and activation computes remain linear to the number of neurons $N/x$. Message traffic of $l_i$, which we consider across all neurocores, is $mxN$ messages duplicated to $x^2$ neurocores in $l_{i+1}$. Overall:
\begin{enumerate}[label=(\alph*)]
    \item Synops per neurocore: $O(mwN^2)$
    \item Act. computes per neurocore: $\sim O(N / x)$
    \item Message traffic to $l_{i+1}$: $O(mx^3N)$
\end{enumerate}

Thus, \textbf{increasing network width forces a quadratic increase in neurocore utilization, and a cubic increase in message traffic volume.} In addition, if each neurocore is filled to its capacity (i.e., no `voluntary' partitioning is applied on top of `involuntary' utilization), \textbf{synops per neurocore will not change as a function of network width.} Together, this suggests that wide network layers will be likely to cause NoC traffic to overtake the previously-identified dominant bottleneck of neurocore synops.

\subsection{Key Modeling Insights}
We summarize the findings into key Modeling insights (M), presenting guidelines to understanding the neuromorphic performance bottleneck implications across workload configurations:


\textbf{M1.} \textit{Memory-bound}—Synops are a dominant bottleneck, scaling quadratically in network width until saturating the capacity of a neurocore. This bottleneck 
can be addressed by increasing weight and activation sparsity, or increasing partitioning. We refer to this as memory-bound since synop costs are mainly weight fetch and sum writeback~\cite{Furber_2016, tang2023openboxdigitalneuromorphic}.

\textbf{M2.} \textit{Compute-bound}—Activation computes are unlikely to bottleneck performance since they only scale linearly with the number of neurons in a neurocore. However, if sparsity and partitioning is large enough to shrink the synop bottleneck without triggering a traffic bottleneck, the workload may be compute-bound.

\textbf{M3.} \textit{Traffic-bound}—High neurocore utilization from partitioning or forced from large network widths will expand the traffic volume linearly or cubically, leading to a traffic-bound state. But, an intelligent mapping of neurocores is also useful for mitigating the traffic bottleneck, i.e., by placing high-output neurocores on separate NoC router paths.

Moreover, we emphasize another key insight of our analytical model: synop and compute bottlenecks appear at the neurocore level, not the total network level. Therefore, total weight sparsity and activation sparsity will be inaccurate performance indicators if the resulting synops are unevenly distributed across neurocores, particularly for barrier-synchronized neuromorphic accelerators where total performance is dependent on the slowest neurocore.
Aligning with the enumeration, we refer to this as \textbf{M0.}

In Section~\ref{sec:characterization}, we substantiate each of these insights by extensively profiling real neuromorphic accelerators, which we briefly introduce next.

\section{Accelerators and Workloads}
\label{sec:systems_workloads}


\begin{table}[h!]
\centering
\begin{tabular}{|l|l|l|l|}
\hline
\textbf{Accelerator} & \textbf{Task} & \textbf{Neuron Type} & \makecell[l]{\textbf{Network}\\\textbf{Architecture}} \\
\hline
AKD1000~\cite{brainchipakd1000} & \makecell[l]{RGB image\\classification~\cite{imagenette}} & ReLU & CNN~\cite{howard2017mobilenetsefficientconvolutionalneural} \\
\hline
Speck~\cite{man24speck} & \makecell[l]{Event camera\\classification~\cite{orchard2015converting}} & Spiking IF & CNN \\
\hline
\multirow[t]{2}{*}{Loihi 2~\cite{loihi2techbrief}} & \makecell[l]{RGB video\\regression~\cite{bojarski2016endtoend}} & SD-ReLU~\cite{o'connor2017sigma} & CNN~\cite{shrestha2024efficient} \\
\cline{2-4}
 & \makecell[l]{Audio\\denoising~\cite{Timcheck_2023}} & ReLU & SSM~\cite{smith2023simplified} \\
\hline
\end{tabular}
\caption{Summary of accelerators and workloads under study.}
\label{tab:system_summary}
\end{table}

Our cross-platform empirical investigation spans three neuromorphic accelerators with fundamentally different architectural implementations and target applications, allowing us to establish performance principles that generalize across neuromorphic designs rather than being specific to individual implementations. The platforms represent distinct points in the neuromorphic design space: edge inference acceleration (AKD1000), specialized sensor processing (Speck), and research programmability (Loihi 2). 
Table~\ref{tab:system_summary} summarizes the three accelerators and four workloads under study.


\subsubsection{AKD1000} 
The Brainchip AKD1000~\cite{brainchipakd1000} is an edge inference accelerator with 80 neurocores optimized for ReLU activation sparsity in CNNs. This platform enables investigation of CNN weight sparsity limitations and load balancing effects. We characterize this accelerator using AkidaNet~\cite{akidanet}, a variant of the depthwise-separable MobileNetv1 CNN~\cite{howard2017mobilenetsefficientconvolutionalneural} using standard convolutions in early layers. The network is applied to classify Imagenette~\cite{imagenette}, a 10 class subset of the ImageNet dataset ~\cite{imagenet}


\subsubsection{Speck}
The Synsense Speck~\cite{man24speck} is a specialized neuromorphic architecture with 9 neurocores optimized for event camera~\cite{indiveri00dvs} processing using spiking integrate and fire neurons. In this study, this platform presents a highly specialized, micro-edge accelerator, which we demonstrate still conforms to the general trends as the other more general-purpose platforms.
Unlike barrier synchronized accelerators (AKD1000 and Loihi~2), Speck neurocores operate asynchronously and enter idle states when no input is present~\cite{richter2024speck}. Furthermore, there is no partitioning in network layers—each neurocore maps a full layer.


\subsubsection{Loihi 2}
The Intel Loihi~2~\cite{loihi2techbrief} is our primary characterization platform, containing 120 programmable neurocores and having extensive software instrumentation for micro-architectural analysis. The platform enables arbitrary network partitioning, detailed neurocore operation analysis, and validation of performance scaling principles across diverse network architectures and utilization regimes.
We study two representative workloads:
First, PilotNet~\cite{shrestha2024efficient}, a CNN of sigma-delta ReLU (SD-ReLU) neurons for processing RGB video sequences into vehicle steering angle~\cite{bojarski2016endtoend}. The sigma-delta neurons~\cite{o'connor2017sigma} leverage the high ReLU similarity between consecutive video frames to produce high activation (message) sparsity on-chip.
Second, S5, a state space model~\cite{smith2023simplifiedstatespacelayers} adapted for activation sparsity on Loihi~2 by adding ReLU activations~\cite{pierro2025acceleratinglinearrecurrentneural}, applied to an audio denoising task~\cite{Timcheck_2023}. To avoid I/O latency bottlenecks~\cite{shrestha2024efficient}, we study accelerator performance using input data loaded into neurocores on chip.

\section{System Characterization}
\label{sec:characterization}

We substantiate the theoretical modeling insights from Section~\ref{sec:modeling} using a broad empirical characterization of the real neuromorphic accelerators. We demonstrate the validity of the insights, in that each of the bottlenecks do appear under configurations expected by the model. Furthermore, we profile the relative performance impacts of each bottleneck on overall accelerator performance, extending the asymptotic trends to real performance boundary curves.

\subsection{Characterization Methodology}

By manually configuring network weights, biases, and thresholds, or programming neurons directly (i.e., non-trained networks), we systematically profile the full range of workload sparsity and execution dynamics that a trained network can take.

Our profiling is structured at two levels of granularity. First, we examine network-wide weight sparsity and activation sparsity for all three accelerators, to investigate the shortcomings of total, high-level metrics (\textbf{M0}). Then, using the Loihi~2, we conduct a deeper dive into investigating workload configurations that exhibit the three bottleneck states of memory (\textbf{M1}), compute (\textbf{M2}), and traffic (\textbf{M3}) that are suggested by the analytical model.

We use two performance metrics: \textbf{Time per Step} measures the average duration between consecutive inference outputs, corresponding to timestep duration for synchronized accelerators (AKD1000, Loihi~2), and full sample processing time for the fully asynchronous Speck. \textbf{Energy per Step} combines average system power with execution time to evaluate efficiency. All measurements directly use the accelerators' telemetry and software instrumentation, except for the timing of Speck, which is measured in wall-clock time over a long experiment. \textbf{In our empirical study, we normalize all performance results and micro-operation measurements to focus on performance bound and bottleneck trends, rather than absolute system comparisons.}

\subsection{\textbf{M0:} Network-wide Weight Sparsity}

\begin{figure}[h]
  \centering
  \begin{subfigure}{0.24\textwidth}
    \centering
    \includegraphics[width=\textwidth]{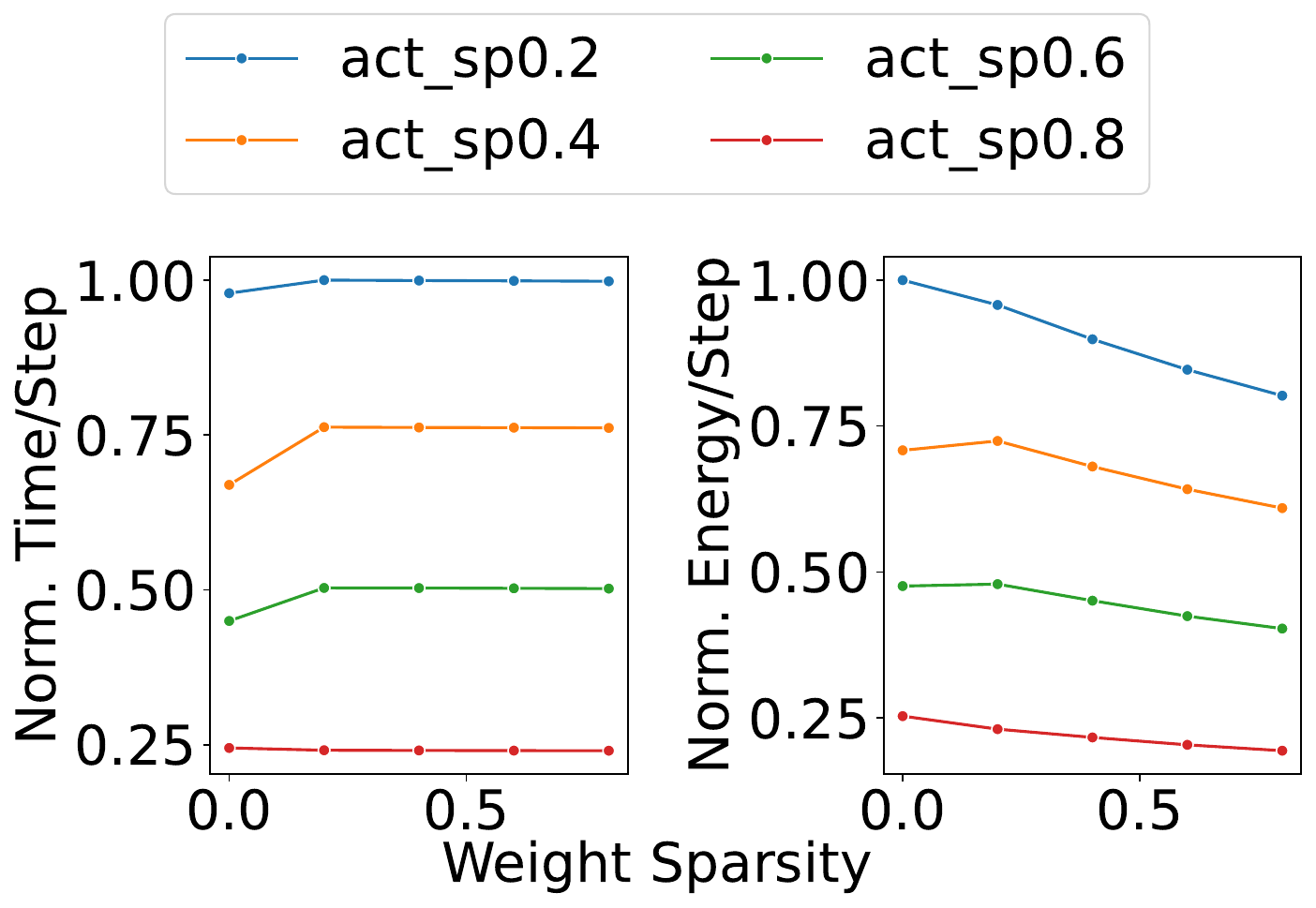}
    \caption{AKD1000}
  \end{subfigure}
  \begin{subfigure}{0.24\textwidth}
    \centering
    \includegraphics[width=\textwidth]{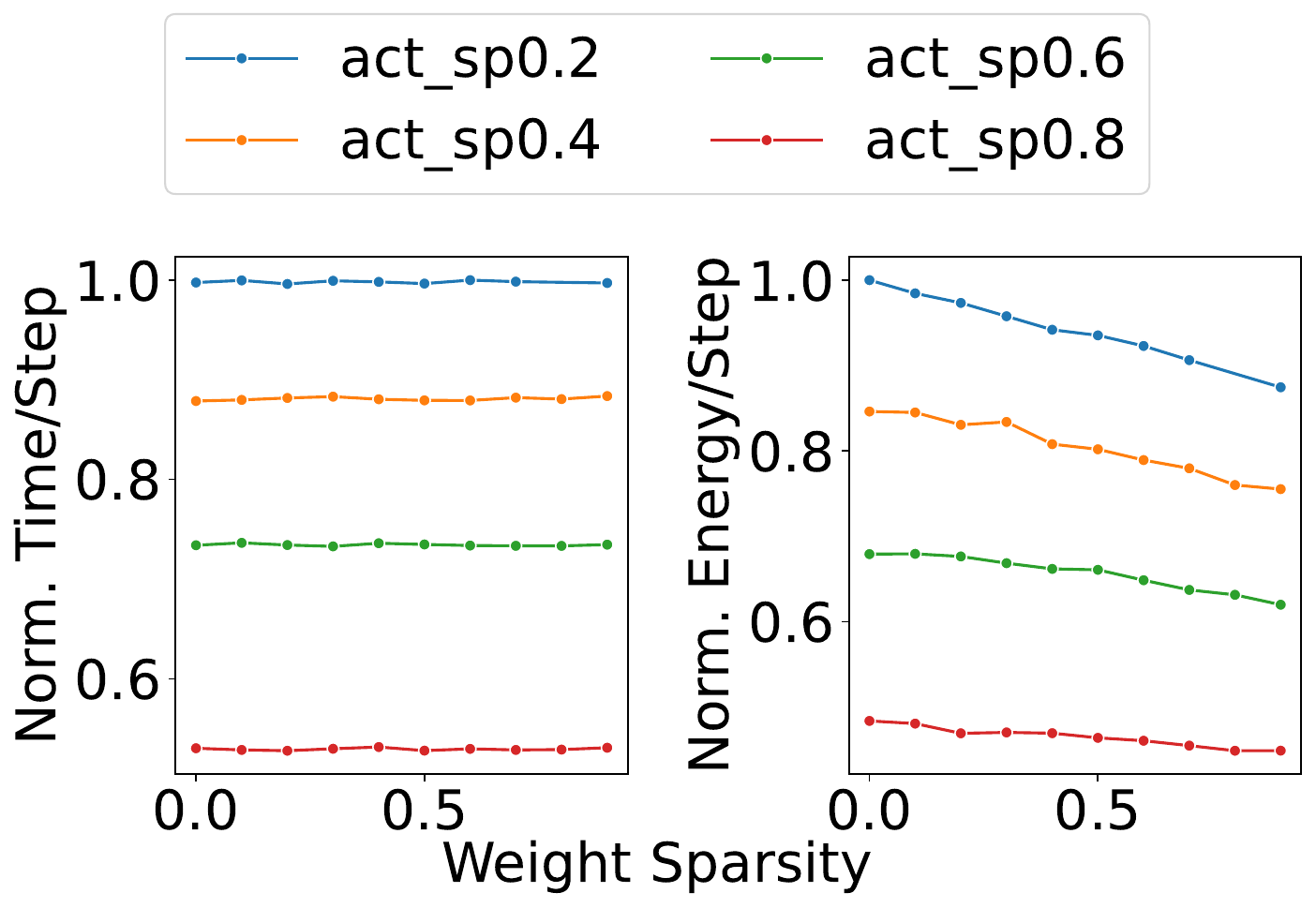}
    \caption{PilotNet (Loihi 2)}
    \end{subfigure} \\
    \vspace{1em}

  \begin{subfigure}{0.48\textwidth}
    \centering
    \includegraphics[width=.4\textwidth]{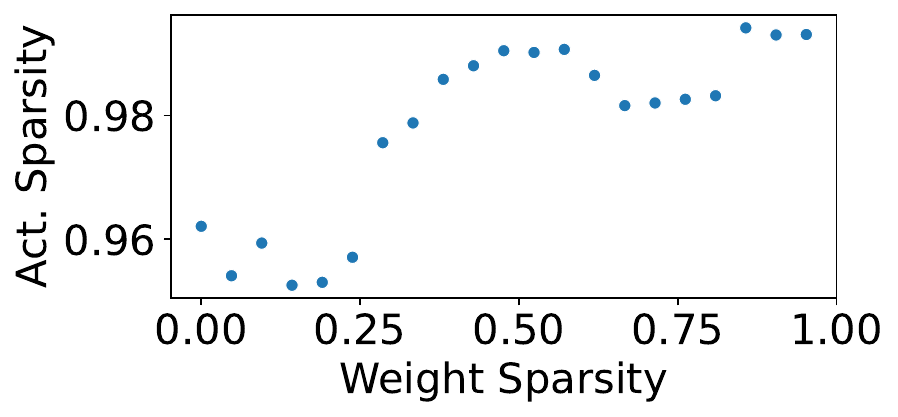}
    \includegraphics[width=.58\textwidth]{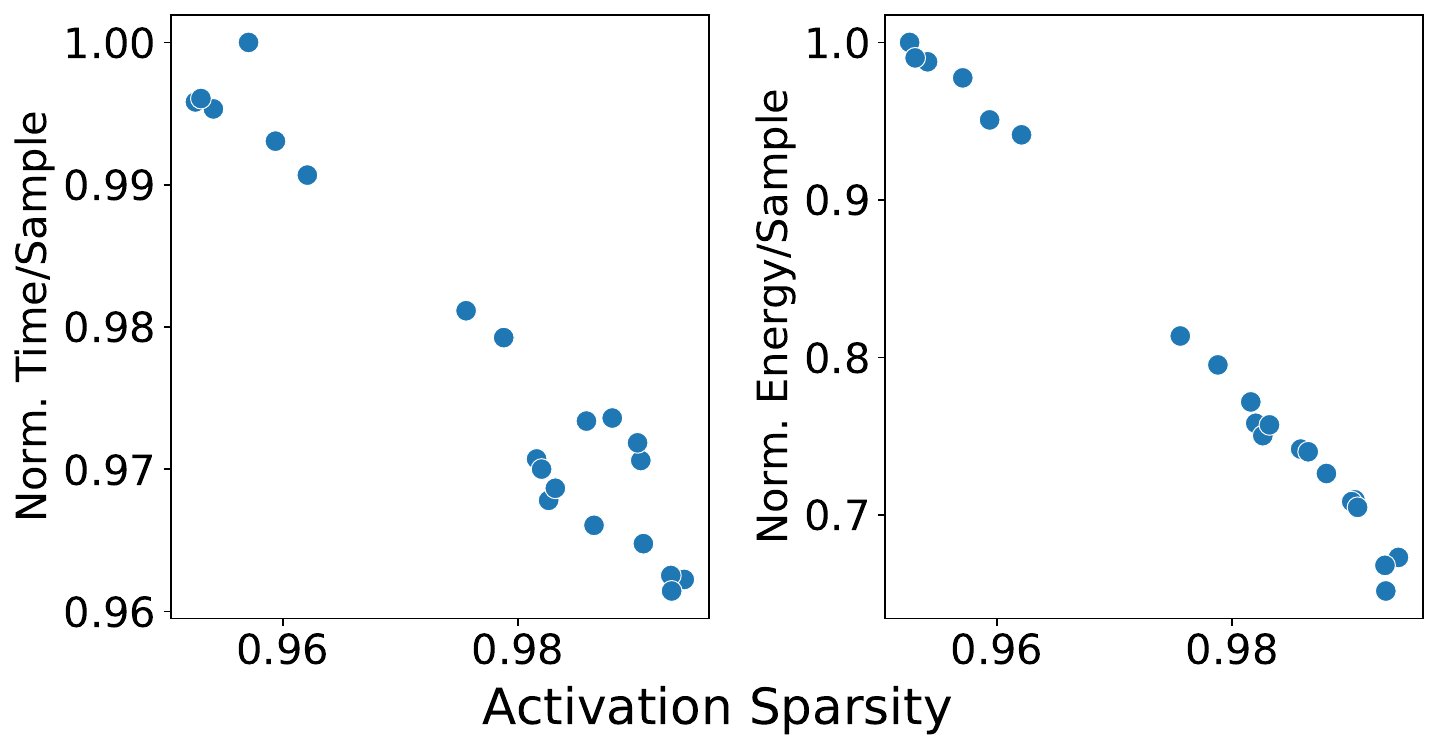}
    \caption{Speck}
  \end{subfigure}\\
  \caption{Weight sparsity performance of CNNs.}
  \label{fig:weight_sparsity_cnn}
\end{figure}

\begin{figure}[h]
    \centering
    \includegraphics[width=0.40\textwidth]{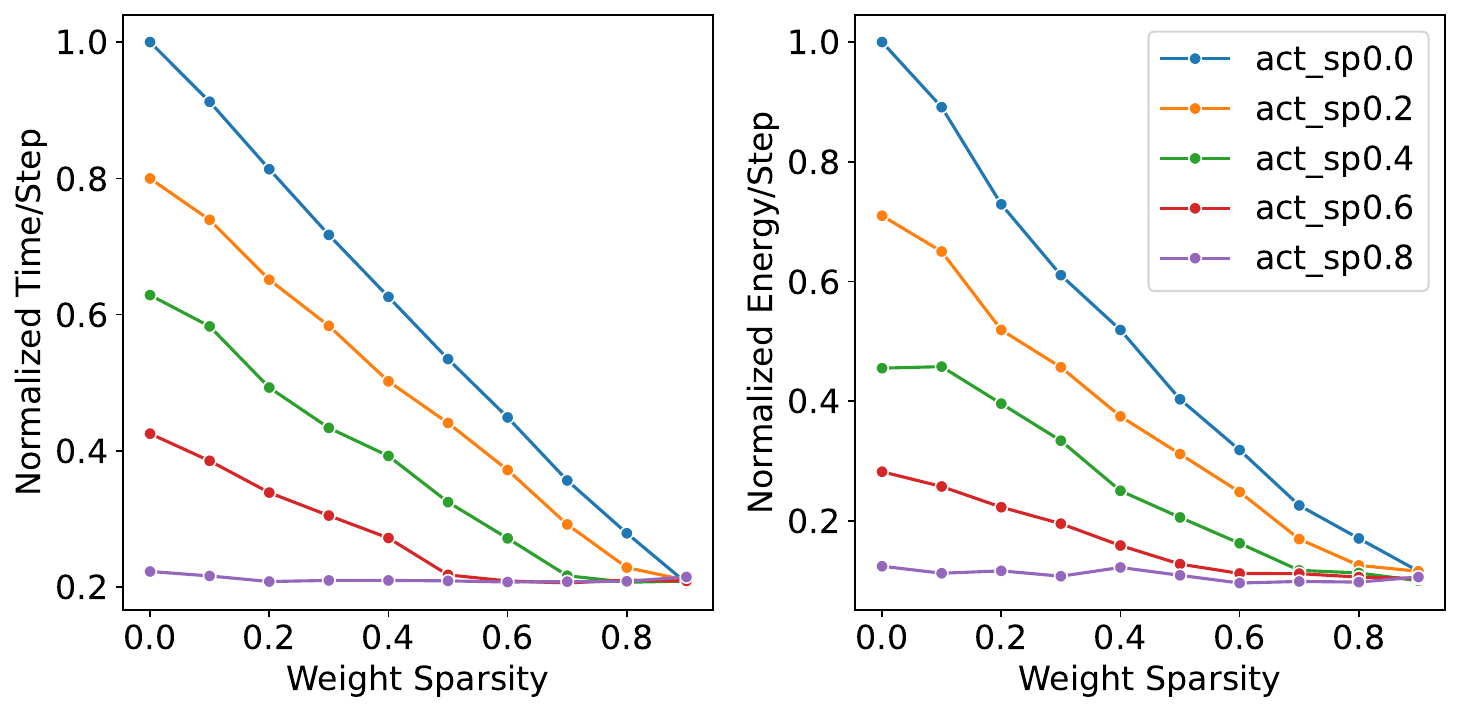}
    \caption{S5 (Loihi 2) weight sparsity performance}
    \label{fig:weight_sparsity_s5}
\end{figure}

We begin by studying whether total weight sparsity is an accurate performance indicator (\textbf{M0}), and surprisingly observe that for CNN workloads, weight sparsity does not yield substantial performance benefits.

Figures~\ref{fig:weight_sparsity_cnn} and~\ref{fig:weight_sparsity_s5} profile the effects of weight sparsity on the timing and energy performance, by tracking uniformly increasing weight sparsity at a constant activation sparsity (act\_sp). Since Speck cannot directly control for activation sparsity, we show the effect of increasing weight sparsity on activations, then re-visualize these networks with performance as a function of their activation sparsity.

In Figure~\ref{fig:weight_sparsity_cnn}, for the CNNs on AKD1000 and Loihi 2, there is no benefit for runtime, and a slight benefit for energy, likely due to compute efficiency for zero values. 
The magnitude of energy improvement is small, for instance, the energy benefit from increasing weight sparsity from 0.0 to 0.9 is smaller than the benefit of increasing activation sparsity from 0.2 to 0.4.
Similarly, on the Speck, energy performance remains linear to activation sparsity, and there is no correlated performance benefit with increasing weight sparsity. 

In contrast, Figure~\ref{fig:weight_sparsity_s5} shows that weight sparsity is supported for the linearly connected S5 network, and its benefit is roughly equal to the benefit of increasing activation sparsity. For example, the performance benefit by increasing weight sparsity from 0.0 to 0.2 with 0.0 activation sparsity (blue line), is about equal to the improvement of increasing activation sparsity from 0.0 to 0.2 with 0.0 weight sparsity (blue to orange line).

\begin{figure}[h]
    \centering
    \includegraphics[width=0.40\textwidth]{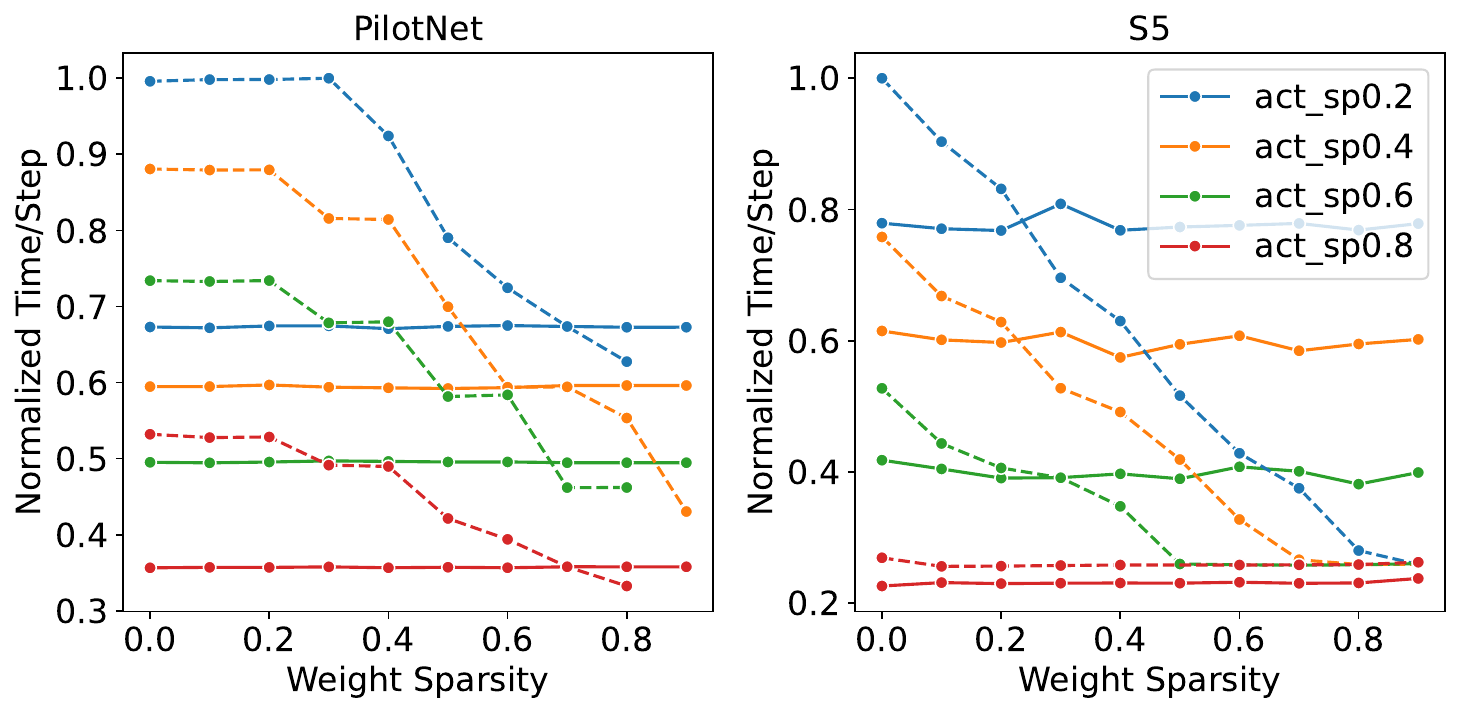}
    \caption{Timing overhead of sparse weight support on Loihi 2. Solid lines use dense weight formatting, while dashed lines use sparse weight formatting.}
    \label{fig:dense_vs_sparse}
\end{figure}

The weight sparsity results in Figures~\ref{fig:weight_sparsity_cnn} and~\ref{fig:weight_sparsity_s5} use the default weight formatting on Loihi~2: dense for CNNs, sparse for linearly connected networks.
Figure~\ref{fig:dense_vs_sparse} shows the implications of sparse weight formatting for PilotNet and S5. For the PilotNet CNN, the sparse weight format only begins to improve timing performance around 0.7 weight sparsity, under which the timing performance is much worse. For the S5 linearly connected network, the sparse weight format is beneficial around 0.2 weight sparsity. In the remainder of the paper, we use the default formatting.

The challenge of supporting weight sparsity in CNNs using the neuromorphic execution paradigm is that kernel weight fetches triggered for each input activation are small, so the overhead of decoding the sparse weight format can outweigh the benefit of fewer weight memory accesses. The profiling implies that specialized architecture components, potentially utilizing structured weight sparsity, may be necessary to support CNN weight pruning methods which have been intended for neuromorphic deployment optimization~\cite{shi2024towards, shen23esl-snn, chakraborty2024sparse}.

\subsection{\textbf{M0.} Network-wide Activation Sparsity}

\begin{figure}[t]
    \centering
    \includegraphics[width=0.48\textwidth]{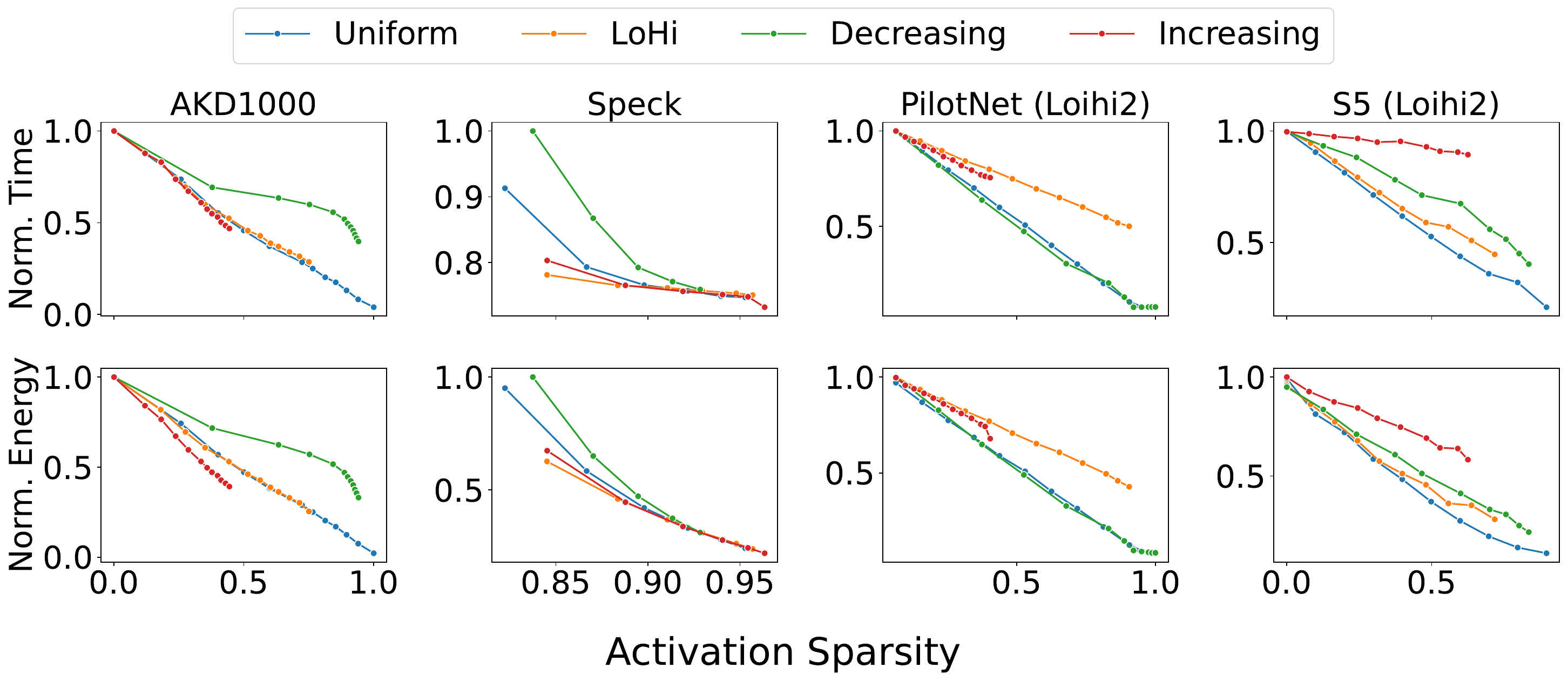}
    \caption{Activation sparsity with varying sparsity schedules to change load balance}
    \label{fig:act_sparsity_schedules}
\end{figure}

Next, we repeat the investigation on total activation sparsity, particularly under load imbalanced conditions (\textbf{M0}).

Figure~\ref{fig:act_sparsity_schedules} shows the relationship between activation sparsity and performance. 
In this characterization, we skew load balance across neurocores by using different sparsity schedules: Uniform is consistent sparsity across all layers, LoHi is an alternating scheme of low and high sparsity, and Decreasing and Increasing linearly scale sparsity from first to last layer. 

In the AKD1000 and Loihi~2 workloads, the sparsity is exactly programmed by explicitly toggling neuron activation messaging on and off. These workloads exhibit a linear correlation between sparsity and performance when the sparsity is Uniform across network layers. However, when the sparsity is non-uniformly applied in the LoHi, Decreasing, and Increasing schedules, the workload is no longer balanced, and the correlation breaks down. 
In the Speck, the sparsity schedule is programmed as spiking neuron thresholds per network layer, as exact sparsity cannot be induced. The LoHi and Increasing schedule curves have better performance at the same total sparsity, suggesting that the last network layer (which has the most weights) is the workload bottleneck.

In effect, we confirm that total network measurements like activation sparsity can be correlated with performance benefits, but they are insufficient for fully understanding performance due to the parallelization of the network across neurocores.




\subsection{\textbf{M1.} Dominant Memory Bottleneck of Max SynOps}

Next, we profile the three bottleneck states suggested by the analytical modeling in Section~\ref{sec:modeling}, using a deeper dive into Loihi 2. We begin with the expected dominant bottleneck, neurocore synops (\textbf{M1}).

\begin{figure}[h]
  \centering

  \begin{subfigure}{0.40\textwidth}
    \centering
    \includegraphics[width=\textwidth]{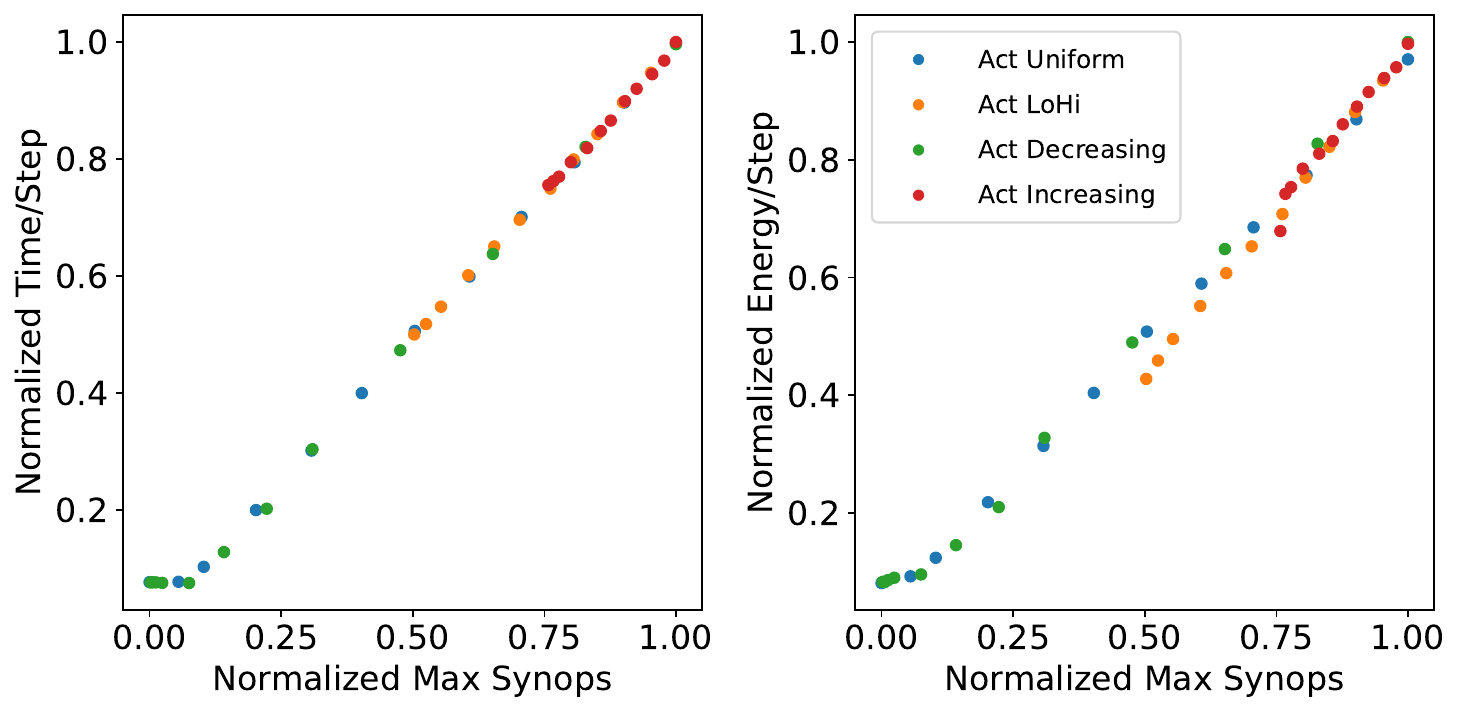}
    \caption{PilotNet}
    \label{fig:mem_bound_pilotnet}
  \end{subfigure}\\
  \vspace{1em}

  \begin{subfigure}{0.40\textwidth}
    \centering
    \includegraphics[width=\textwidth]{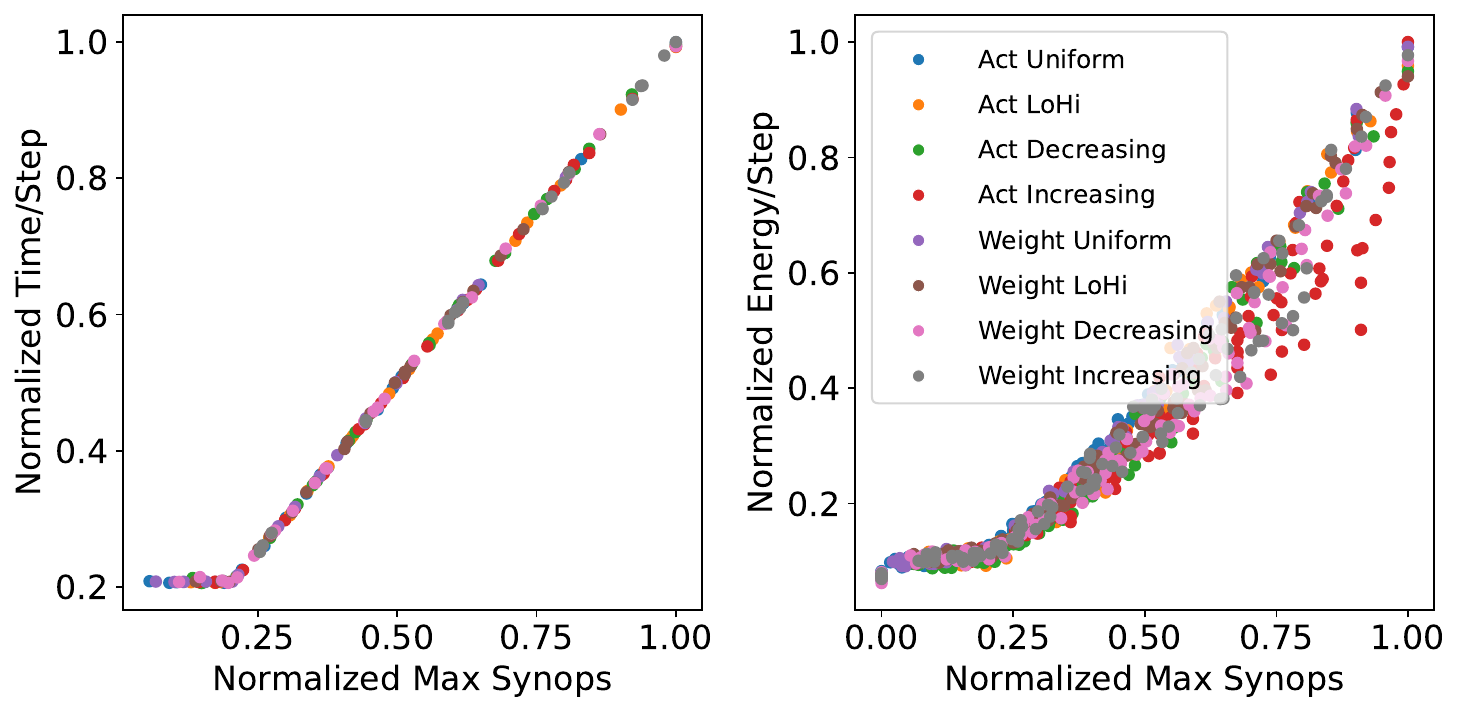}
    \caption{S5}
    \label{fig:mem_bound_s5}
  \end{subfigure}\\

  \caption{Memory-bound bottleneck for all sparsity schedules of PilotNet and S5.}
  \label{fig:mem_bound}
\end{figure}

Figure~\ref{fig:mem_bound} shows all PilotNet and S5 activation sparsity schedule configurations from Figure~\ref{fig:act_sparsity_schedules}, plotted with performance as a function of the maximum synops of any active neurocore. Since the S5 network can also leverage weight sparsity (Figure~\ref{fig:weight_sparsity_s5}), we additionally include S5 workloads with weight sparsity applied in imbalanced sparsity schedules. Here, all workloads for the same network architecture use the same, minimal neurocore partitioning.

Despite widely varying sparsity and load balance configuration, the timing performance of Loihi 2 is linear to the max synops metric, until reaching a performance floor when the max synops is sufficiently low. Energy, as well, is well-correlated with this metric. This profile not only verifies that max synops presents a performance bottleneck for Loihi 2, but it also demonstrates that this bottleneck results in a clear performance boundary: workload timing does not run faster than the linear max synops limit.



\subsection{\textbf{M2.} Compute Bottleneck Floor with Low Max Synops}

\begin{figure}[t]
    \centering

    \begin{subfigure}{0.40\textwidth}
    \centering
    \includegraphics[width=\textwidth]{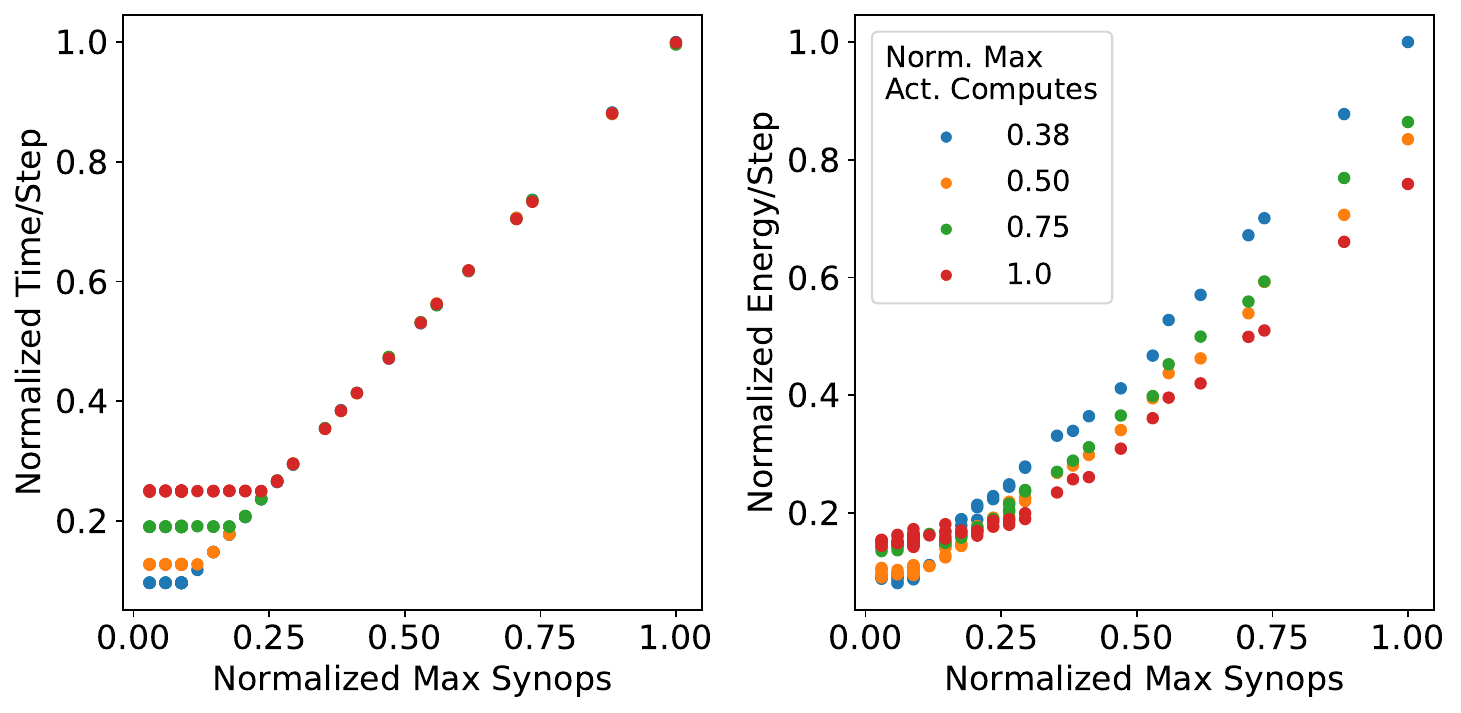}
    \caption{PilotNet}
    \label{fig:compute_bound_pilotnet}
    \end{subfigure}\\
    \vspace{1em}
    
    \begin{subfigure}{0.40\textwidth}
    \centering
    \includegraphics[width=\textwidth]{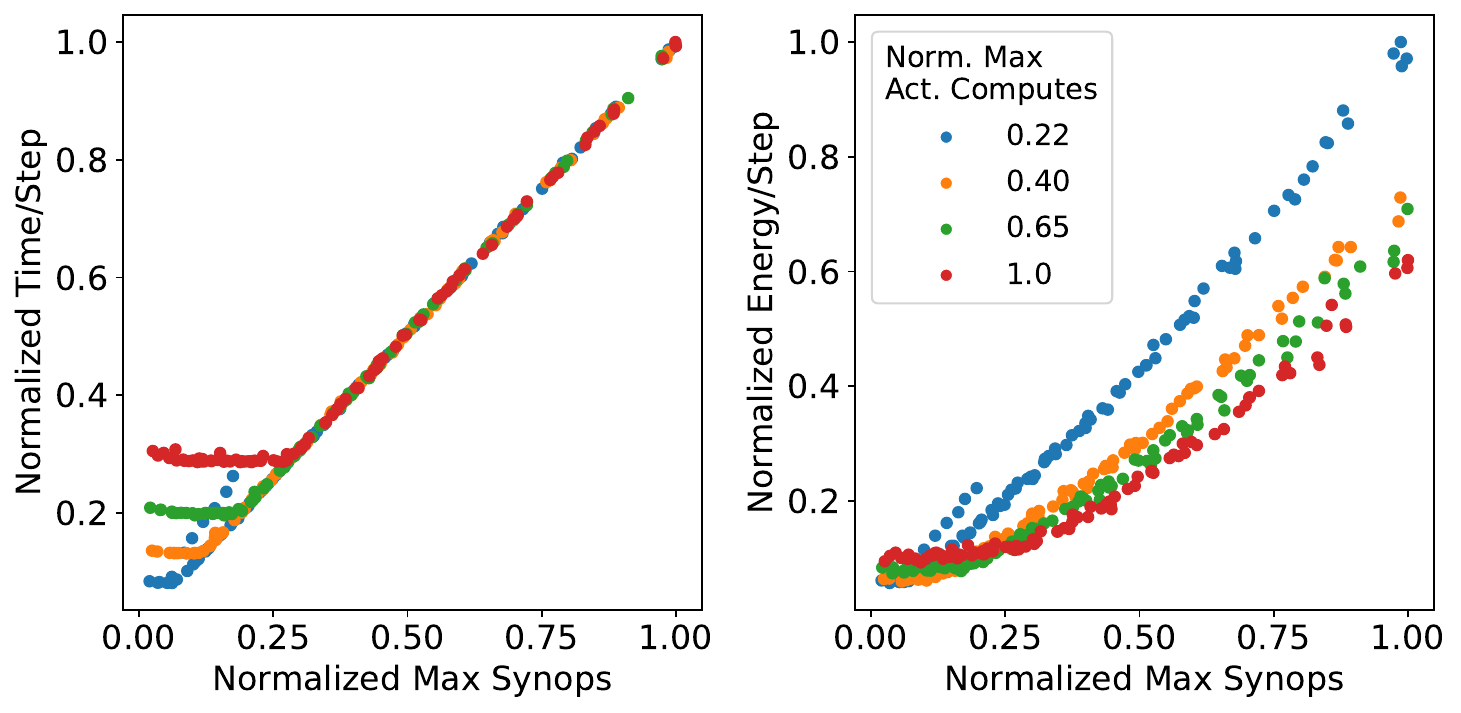}
    \caption{S5}
    \label{fig:compute_bound_s5}
    \end{subfigure}\\
    
    \caption{Lowering the compute-bound floor by increasing partitioning.}.    
    \label{fig:compute_bound}
\end{figure}


The timing floor that appears at low max synops is due to the workloads shifting from a memory-bound synops bottleneck state to an activation computation bottleneck state (\textbf{M2}). In Figure~\ref{fig:compute_bound} (left), we verify this for PilotNet and S5 by demonstrating that the timing floor is lowered when the layer with the max activation computes is partitioned, which reduces the number of neurons in the compute-bottleneck neurocore and matches the expected behavior from Section~\ref{sec:voluntary_utilization}. Thus, the compute-bound timing floor is a variable performance boundary, dependent on the max activation computes of the workload partitioning configuration.

Figure~\ref{fig:compute_bound} (right) also shows that energy curves diverge among the partitioning settings. As partitioning increases, lowering the max activation computes, more neurocores are utilized, raising power and thus raising energy as well. In Figure~\ref{fig:compute_bound}, the PilotNet utilization increases from 80 to 119 neurocores (highest to lowest timing floor). For the S5, the utilization increases from 27 to 60 neurocores.
In general, the energy result implies that unless timestep duration is being improved by lowering the compute floor, an optimal network configuration should use the fewest neurocores possible, as adding cores leads to greater power.

\subsection{\textbf{M3.} Traffic Bottleneck with High Neurocore Utilization}

\begin{figure}[t]
    \centering

    \begin{subfigure}{0.40\textwidth}
    \centering
    \includegraphics[width=\textwidth]{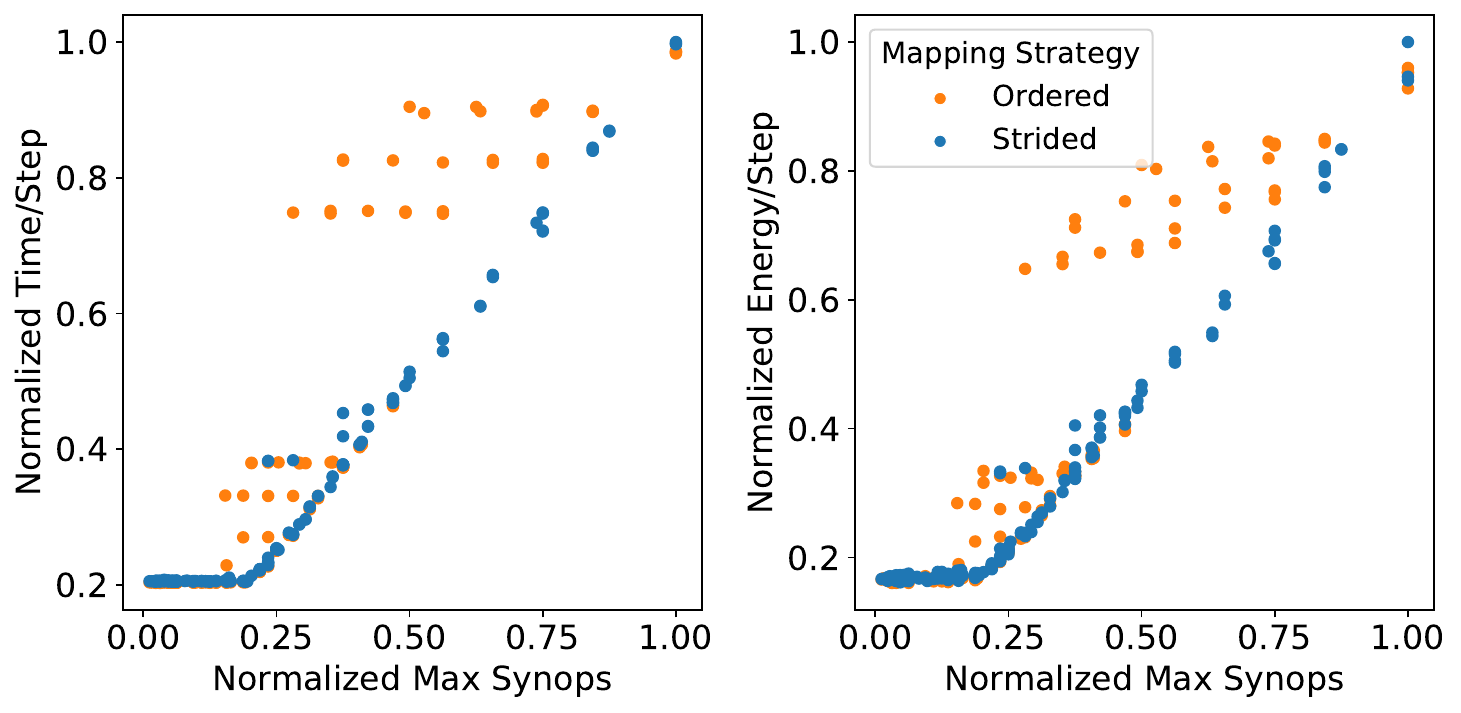}
    \caption{PilotNet}
    \label{fig:traffic_bound_pilotnet}
    \end{subfigure}\\
    \vspace{1em}
    
    \begin{subfigure}{0.40\textwidth}
    \centering
    \includegraphics[width=\textwidth]{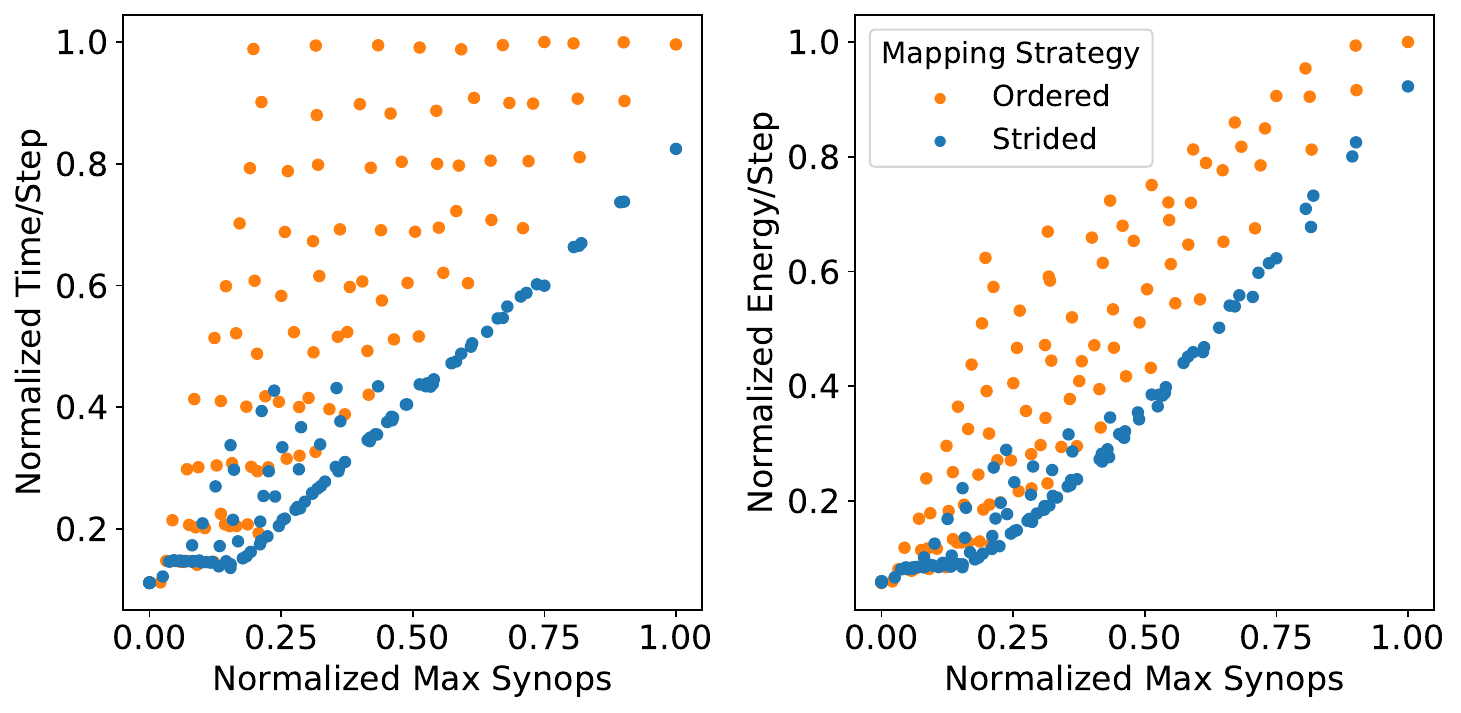}
    \caption{S5}
    \label{fig:traffic_bound_s5}
    \end{subfigure}\\
    
    \caption{Traffic-bound comparison between ordered mapping (orange) and strided mapping (blue), using high neurocore utilization workloads.}. 
    \label{fig:traffic_bound}
\end{figure}

We address the third bottleneck state, traffic-bound (\textbf{M3}), by profiling workloads with high neurocore utilization.

For PilotNet, we reduce the convolution spatial dimensions and channels, shrinking the network to free more neurocores for greater (voluntary) partitioning (Section~\ref{sec:voluntary_utilization}). 
Scaling the network width does not significantly affect CNN utilization, as neurons maintain small receptive fields.
The reduced PilotNet configuration when minimally partitioned uses 32 neurocores, and the network was partitioned to 116 neurocores. For reference, the prior full-size PilotNet profiling configurations use their minimum utilization of 96 neurocores.

For S5, we double the network width in order to increase (involuntary) utilization (Section~\ref{sec:involuntary_utilization}). This configuration uses 76 neurocores, compared to the prior S5 profiling configurations which use 27 neurocores.

With the high-utilization network configurations, and still sweeping over sparsity configurations, Figure~\ref{fig:traffic_bound} shows the difference between a poor ordered mapping and a better strided mapping. The ordered mapping places neurocores sequentially on the chip, a scheme which has been previously used for Loihi~1~\cite{lin18loihimapping}. These workloads struggle with a traffic bottleneck since the highest output neurocores, belonging to the same network layer, are physically close to one another and create congestion on their shared NoC routers.

The improved strided mapping places neurocores in strided intervals, separating neurocores of the same network layers to different NoC routing paths. While the improved heuristic mapping does not fully address the traffic bottleneck for all workload configurations (i.e., strided points still rise above the memory-bound slope and compute floor), it is nevertheless effective at improving performance in all cases, and does not demonstrate negative side-effects such as raising the time or energy floor.

\section{Optimization Approach}
\label{sec:method}

Our performance bound and bottleneck analysis provides valuable insights into how one can improve the performance of a generic neuromorphic accelerator workload. Especially in the current landscape of nascent neuromorphic accelerator research, without standardized micro-architectural designs nor accurate performance estimation models, understanding the critical bottlenecks to look for and how to address these bottlenecks allows for actionable and generalizable performance optimization.

\subsection{The Floorline Performance Model}

\begin{figure}[t]
\centering
\includegraphics[width=0.45\textwidth]{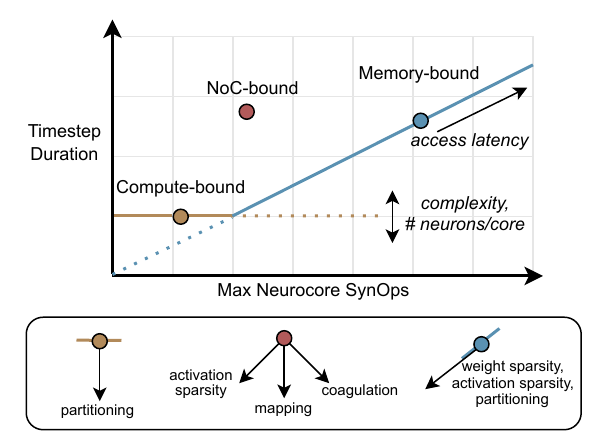}
\caption{The floorline model for understanding performance of neuromorphic systems. Bounds on the timestep duration are defined by memory access latency, or neuron activation computes with sufficiently low memory accesses. Above the bounds, applications are bottlenecked by message traffic in the NoC. 
}
  \label{fig:floorline}
\end{figure}

The insights from the analytical modeling and empirical characterization of real neuromorphic accelerators have identified (1) three bottleneck states and (2) which configurations are likely to exhibit each bottleneck. Moreover, the Loihi~2 characterization shows that the memory and compute bottlenecks result in clear performance boundaries across all sparsity and partitioning configurations for a given network architecture. We distill these insights into a visual model for understanding neuromorphic workload performance and optimization directions: the floorline performance model, shown in Figure~\ref{fig:floorline}. 

In the floorline model plot (Figure~\ref{fig:floorline}, top), the horizontal `intensity' axis is the maximum synops of any active neurocore, and the vertical `performance' axis is timestep duration.
The slope of the floorline is defined by the memory access latency in a neurocore, and represents the memory performance bound of a workload. When the maximum synops decreases, the performance reaches a floor, which is the compute boundary defined by the activation computation. The floor shifts up and down based on the complexity of the bottlenecking activation computation (i.e., number of instructions), and the number of neurons in the compute-bottlenecking neurocore.

The floorline also provides a guide to improving the performance of any neural network workload (Figure~\ref{fig:floorline}, bottom). A trained network, with its distinct sparsity, load balance, and partitioning configuration, will have performance (a) on the slope, (b) on the floor, or (c) above the floorline. These three locations correspond to (a) memory-bound, (b) compute-bound, and (c) traffic-bound states. Optimization of the workload then proceeds as directed by the modeling and characterization insights (\textbf{M1}, \textbf{M2}, \textbf{M3}): 
\begin{enumerate}[label=(\alph*)]
    \item \textit{Memory-bound optimization}—Reduce the maximum neurocore synops by increasing the weight sparsity and/or activation sparsity, or partitioning the synop-bottlenecking neurocore. This will move the workload down and left, improving performance while reducing the synops intensity (max neurocore synops).
    \item \textit{Compute-bound optimization}—Reduce the maximum activation computes by partitioning the compute-bottlenecking neurocore. This will move the workload straight down, improving performance without necessarily affecting the synops intensity.
    \item \textit{Traffic-bound optimization}—Reduce the activation message NoC traffic by increasing activation sparsity or coagulating the network into fewer neurocores. The former moves down and left, potentially reducing synops intensity, while the latter moves down and right, potentially increasing synops intensity but improving performance nonetheless since the workload is traffic-bound. Or, the workload performance can be optimized by improving the neurocore mapping, moving down since the synops intensity will not change.
\end{enumerate}

The floorline differs from the conventional roofline model~\cite{williams09roofline} to capture the unique performance dynamics of neuromorphic accelerators. First, the roofline relates throughput (y-axis) with operational intensity (x-axis) and bandwidth (roofline slope), while the floorline relates timestep duration with operation count and latency. This is suited for the sparse, event-based processing on neuromorphic accelerators. Second, conventionally, only one roofline ceiling is presented, representing the peak compute performance boundary of a system. In the floorline, the location of the compute floor boundary is highly variable based on the number of bottlenecking activation computes. Finally, while the roofline informs performance bounds, deeper micro-architectural and kernel-level insights are usually required to optimize a workload, moving it closer to the performance bounds. With the floorline, the location of a workload on the floorline can completely inform its bottleneck state and how to best optimize it, due to the specialized compute pattern of neuromorphic accelerators for neural network processing.

As for all models, the floorline uses assumptions which may not hold true for all potential neuromorphic accelerator designs. For example, the straight performance boundary lines are only realized if synop, activation compute, and messaging execution are sufficiently overlapped in a pipeline. If these operations were fully sequential, performance would be a function of the sum of the three, rather than a function of the individual bottleneck. 
Moreover, neuromorphic approaches like the Speck~\cite{richter2024speck} and SENECA~\cite{imec2024optimizingseneca} have demonstrated asynchronous inference methods without timestepped neurocore synchronization, which disconnects the floorline model's performance metric of timestep duration. Though, the former is dependent on SNNs, and state of the art ML sequence processing requires timestepped token-by-token execution, while the latter relies on input message sorting which has not been demonstrated with multi-neurocore partitioning in a real accelerator.

\subsection{Optimization Framework}
\label{sec:optimization_framework}

We present an actionable, two-stage optimization procedure for improving the performance of a neuromorphic accelerator workload, involving a sparse-aware training stage, and a floorline-informed partitioning stage.

First, networks should be trained with as high sparsity as application accuracy allows. Techniques which increase sparsity approximately uniformly across network layers are preferable, as they will more likely lead to load-balanced configurations when deployed. Given that overall sparsity improvement is well-studied in ML network training and configuration, this provides a prudent first stage in optimization.

Second, with the network weights locked from training, we use the floorline model as a guide for optimizing the partitioning and mapping of the workload to the chip. Any network architecture may have unique performance boundaries, but one can trace along the boundaries using an iterative, backtracking procedure:

To begin, the network is initialized with the minimum possible neurocore utilization, as well as a good heuristic mapping, such as a strided mapping (Figure~\ref{fig:traffic_bound}). These both increase the likelihood that the initialized network will be memory-bound by max synops. Under this assumption, after executing or simulating the network with sample data, we identify the neurocore which executed the most synops, and increase the partitioning of the layer that this neurocore belongs to. If the memory-bound assumption is correct, then the workload will have traced down the memory slope of the floorline, and we repeat the max synops identification and partitioning. 

If partitioning on the memory-bound assumption is not beneficial, then we backtrack the partitioning, as greater utilization without synops improvement is detrimental for power.

After backtracking, we shift the assumption to compute-bound, partitioning based on the max activation computations. Similarly, we can repeat until it is not beneficial, then backtrack. Finally, we may assume that the workload is NoC-bounded, and optimize mapping by moving the greatest message output neurocores onto separate router paths. In general, we can continuously cycle through each assumption and corresponding optimization step.

The iterative partitioning and mapping process stops when there are no further neurocores available on chip, or when energy starts to worsen due to larger utilization power compared to the timing benefits. The iterative process also stops if none of the three optimizations lead to performance improvements, indicating that the workload has reached its true performance boundary for its sparsity dynamics.

\section{Optimization Results}
\label{sec:results}

To demonstrate the potential speedup that iso-accurate networks can achieve with our optimization framework, we train and optimize a sweep of configurations using the same network architectures and inference tasks as before. Similarly to the characterization, we apply the first step of sparsity optimization across all three accelerators, training sparse networks with iso-accuracy to the baseline. Then, in the second optimization step we use Loihi 2 to further optimize the partitioning and mapping of the iso-accurate networks. 

\subsection{Optimization Methodology}

As performance baselines, we use configurations which rely solely on the innate sparsity of ReLU and spiking activations from accuracy-only training, a strategy which has appeared in recent prior works with the AKD1000~\cite{metatfdocs} and Speck~\cite{richter2024speck}, and Loihi2~\cite{shoesmith2025eventproptrainingefficientneuromorphic}. For PilotNet, we directly apply a prior manually-tuned configuration as the baseline~\cite{shrestha2024efficient}.

For training in optimization stage 1, we use existing network training and quantization frameworks for each workload—MetaTF (AKD1000)~\cite{metatfdocs}, Sinabs (Speck)~\cite{sinabs}, Lava DL (PilotNet)~\cite{lava}, and SparseRNNs—and the following training procedures:
\begin{itemize}
    \item \textbf{AKD1000}—The baseline AkidaNet network was trained with cross entropy loss only. To induce ReLU activation sparsity, we add the T$\ell1$ regularizer \cite{yu2024dualsparsetrainingframework}, which we apply to the pre-trained baseline. All networks were also quantized and fine-tuned.
    \item \textbf{Speck}—The baseline spiking network is also trained with cross entropy loss only. To induce greater spiking sparsity, we add the synops regularization loss~\cite{sorbaro20synop_loss}, training from scratch.
    \item \textbf{PilotNet}—For the PilotNet configurations, the same trained network weights are used, and sigma-delta thresholds are modified to adjust activation sparsity. The baseline is a uniform threshold setting across all network layers from prior work~\cite{shrestha2024efficient}. Our approach assigns thresholds uniquely to each layer in order to reach per-layer sparsity targets, in order to encourage more uniform activation sparsity.
    \item \textbf{S5}—We use weight pruning to sparsify the linearly connected network. With a fully dense baseline model, we prune the smallest 0.1 to 0.9 of weights away in one shot, and fine-tune. Networks include ReLU activations for innate activation sparsity which is not explicitly trained.
\end{itemize}

In optimization stage 2, we take the highest-accuracy networks from sparsity training and apply the backtracking partitioning approach described in Section~\ref{sec:optimization_framework}.

For the S5 networks, due to quantization concerns on Loihi 2, we present deployed performance results with simulated message activity matching the per-layer activity measured during GPU inference.

\subsection{Stage 1: Sparsity Optimization}

\begin{figure}[h]
    \centering
    \includegraphics[width=0.48\textwidth]{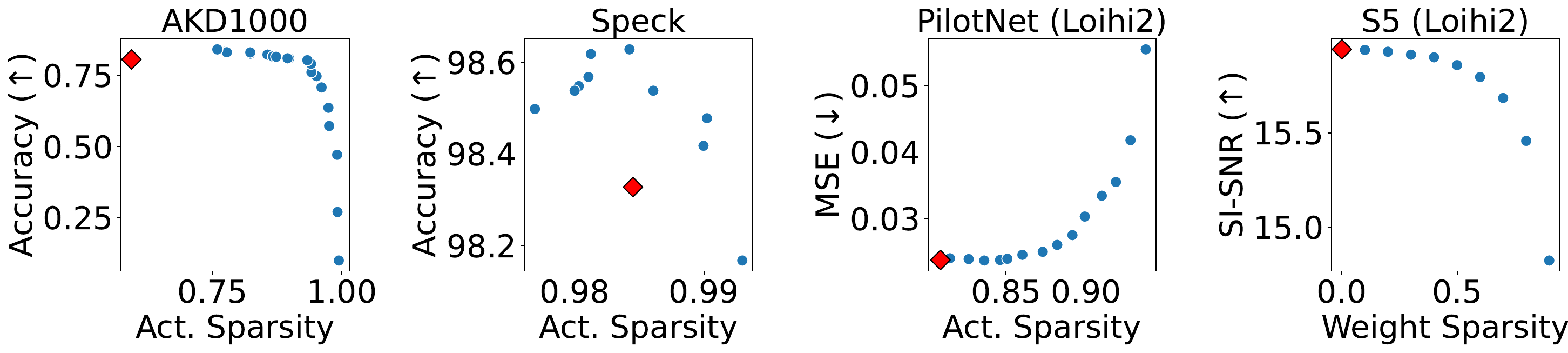}
    \caption{Sparsity vs. accuracy of trained networks. Diamonds indicate the baselines which were not trained with explicit sparsity loss or were presented in prior work (PilotNet~\cite{shrestha2024efficient}).}
    \label{fig:sparsity_accuracy}
\end{figure}

\begin{figure}[h]
    \centering
    \includegraphics[width=0.48\textwidth]{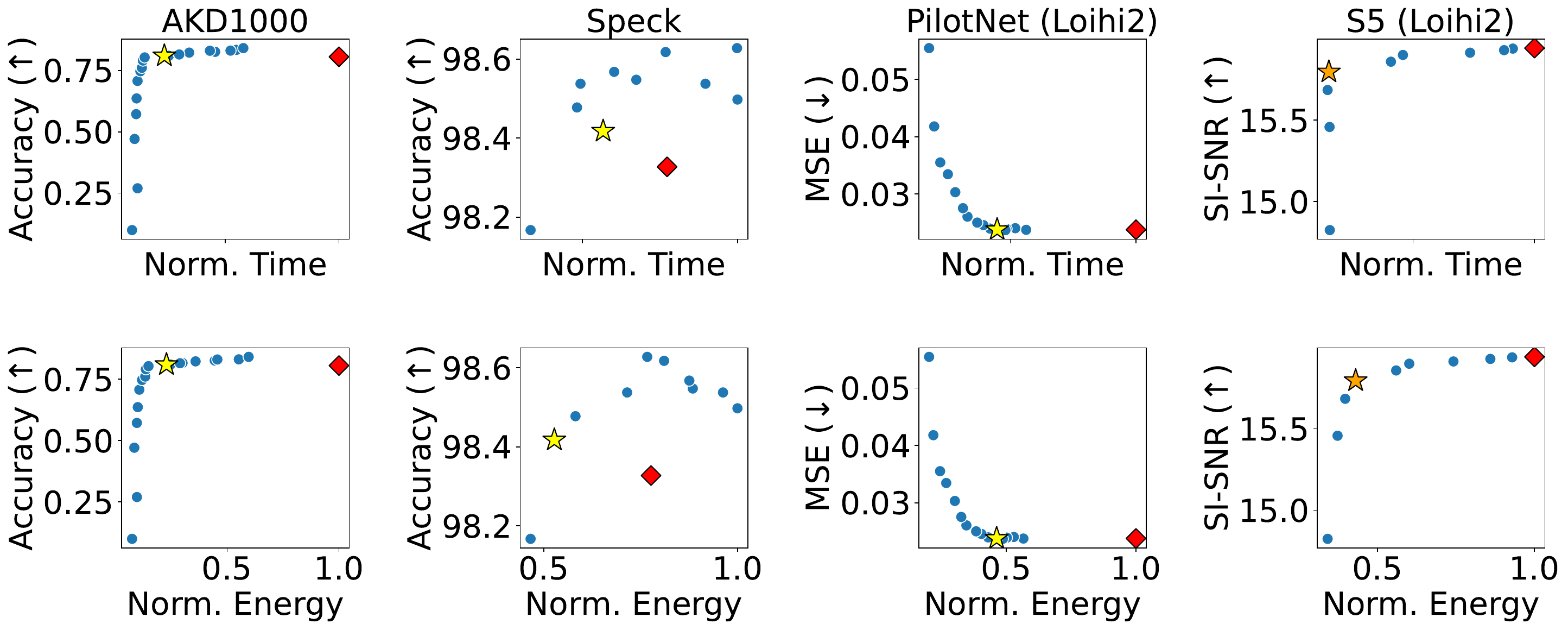}
    \caption{Performance vs. accuracy of the sparse-trained networks and baselines. Diamonds indicate the baselines, while stars mark the highest-performance sparse network configurations which are iso-accurate to the baselines. For S5, the star highlights a high accuracy, high performance network, as no networks are iso-accurate.}
    \label{fig:accuracy_performance}
\end{figure}

Figure~\ref{fig:sparsity_accuracy} shows the accuracy achieved for the range of sparse-trained networks. For the AKD1000, PilotNet, and S5 networks, a clear Pareto curve emerges. For the Speck, which uses spiking neurons, the baseline trained without explicit sparsity regularization achieves considerable spike activation sparsity, but several sparse-trained networks achieve greater accuracy with greater activation sparsity.

Next, Figure~\ref{fig:accuracy_performance} visualizes deployed performance as a function of accuracy, for all of the trained networks, while using constant deployment configurations of partitioning and mapping. Diamonds mark the baselines, and stars mark the highest-performance iso-accurate configurations. Respective time speedup and energy benefit of the iso-accurate workloads against the baselines are: AKD1000 $(4.29\times,4.36\times)$, Speck $(1.01\times, 1.47\times)$, PilotNet $(2.23\times, 2.16\times)$. The pruned S5 networks do not achieve baseline iso-accuracy; though, at a loss of 0.15 dB SI-SNR, the 0.6 sparse pruned network achieves $(1.74\times,2.32\times)$ deployed benefit.

We highlight the large benefit of the uniform sparsity target threshold setting for PilotNet, compared to the baseline's uniform threshold setting. The starred iso-accurate network differs slightly from the baseline in total activation sparsity (1.24$\times$ fewer messages), but our approach achieves much greater deployed performance $(>2\times)$, due to improved neurocore load balance.

\subsection{Stage 2: Partitioning and Mapping Optimization}

\begin{figure}[h]
  \centering

  \begin{subfigure}{0.45\textwidth}
    \centering
    \includegraphics[width=\textwidth]{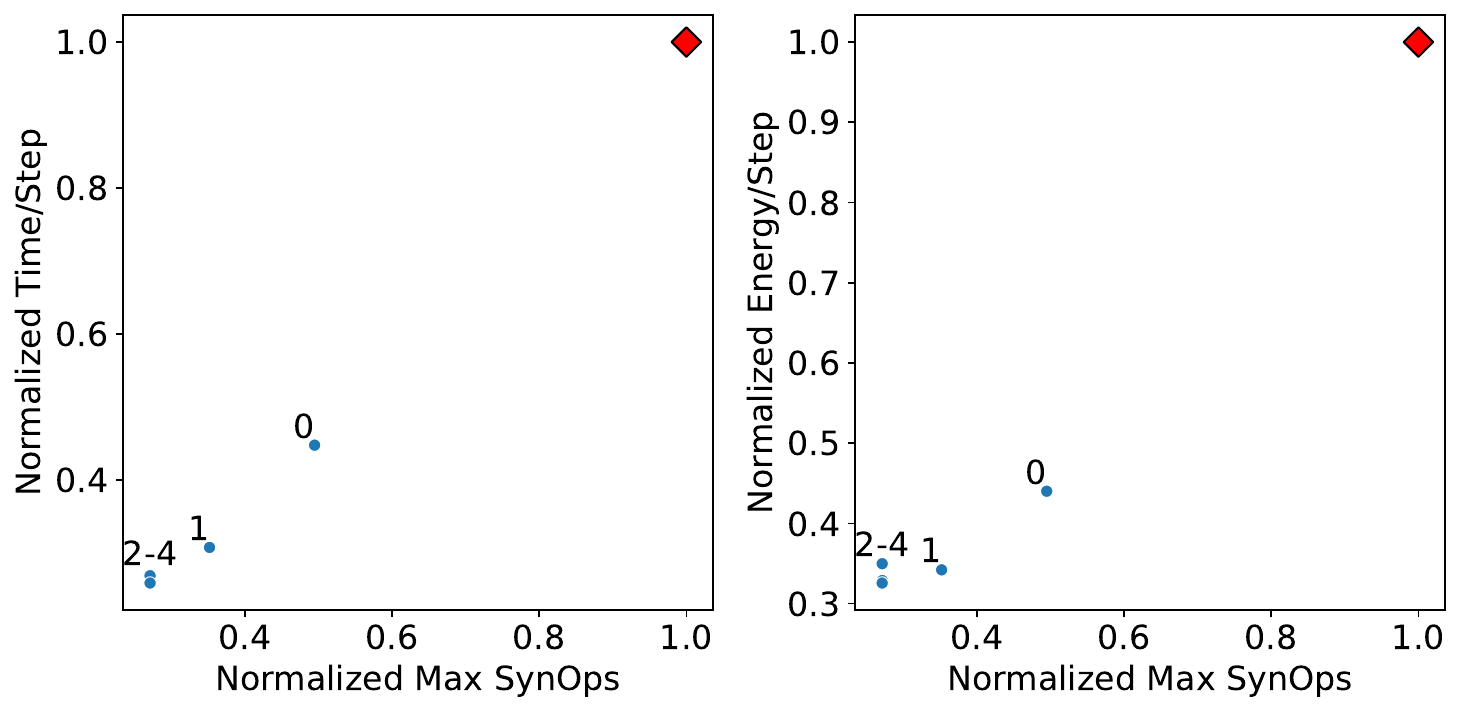}
    \caption{PilotNet}
    \label{fig:partitioning_result_pilotnet}
  \end{subfigure}\\
  \vspace{1em}

  \begin{subfigure}{0.45\textwidth}
    \centering
    \includegraphics[width=\textwidth]{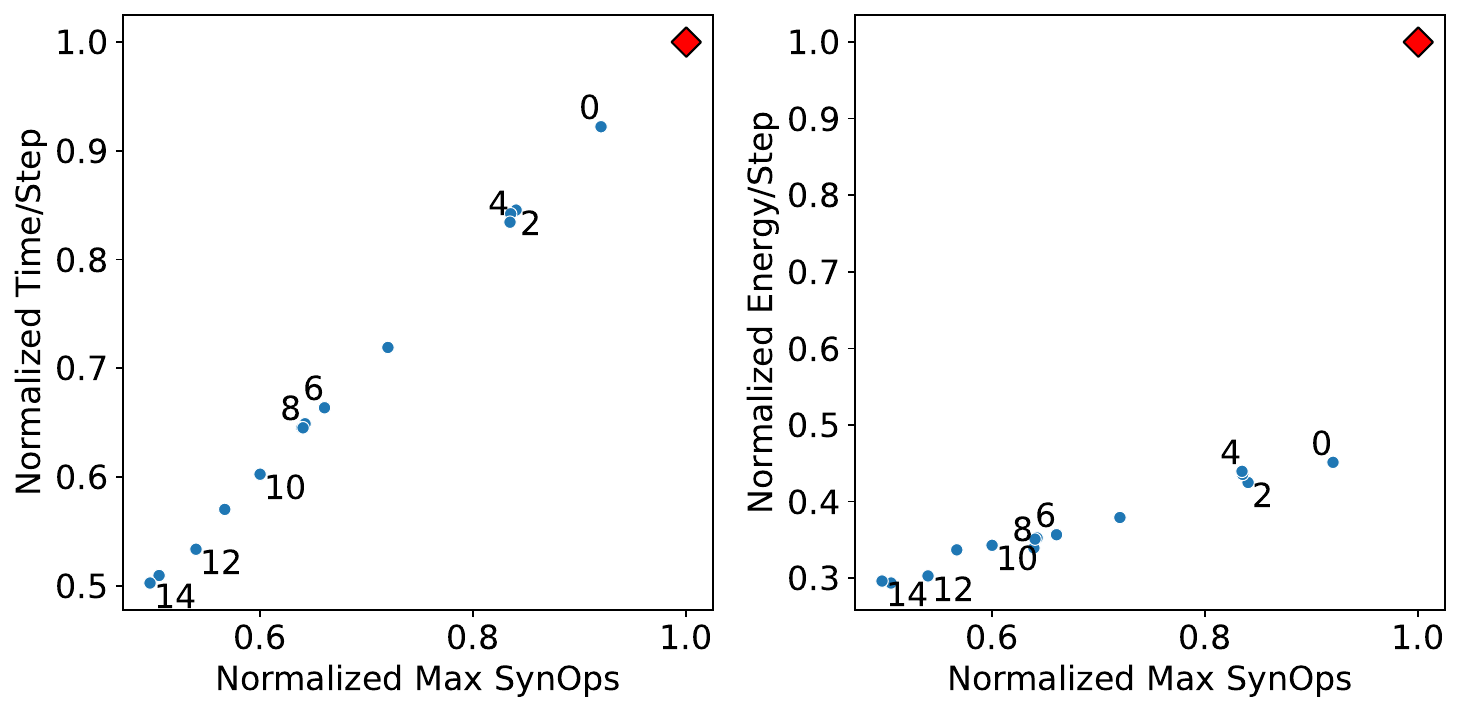}
    \caption{S5}
    \label{fig:partitioning_result_s5}
  \end{subfigure}\\

  \caption{Partitioning and mapping optimization of PilotNet and S5. Diamonds mark the baseline performance from Figure~\ref{fig:accuracy_performance}, which use minimal partitioning. Numbers indicate the partitioning iteration steps, beginning with 0 as the minimally partitioned configuration.}
  \label{fig:partitioning_result}
\end{figure}

Figure~\ref{fig:partitioning_result} shows the further partitioning optimization with the high accuracy sparse star configurations from Figure~\ref{fig:accuracy_performance}. In the figure, the diamonds mark the performance of the baselines (same as Figure~\ref{fig:accuracy_performance} diamonds). For the S5, weight sparsity is used to pack the network into fewer neurocores compared to the dense baseline: in Figure~\ref{fig:partitioning_result_s5}, the minimal partitioning marked with 0 uses 36 neurocores, while the diamond uses 73 neurocores, causing a large utilization power and energy difference.

The performance of the iterative partitioning procedure traces the memory performance boundary for both PilotNet and S5, which means that the optimization was achieved by solely attending to the maximum neurocore synops bottleneck. The PilotNet partitioning ended when no further neurocores were available, while the S5 partitioning ended when energy costs began to rise due to neurocore utilization power without sufficient timing improvements.

The time speedup and energy improvement from only partitioning for the PilotNet is $(1.73\times,1.26\times)$, and for the S5 is $(1.83\times,1.52\times)$. Combined with the sparsity improvements, the total improvements of the final optimized configurations against the baselines are $(3.86\times,2.86\times)$ and $(1.99\times,3.38\times)$ for the PilotNet and S5, respectively. 

\section{Conclusion}
\label{sec:conclusion}

In this work, we presented the first systematic study of performance bounds and bottlenecks in real neuromorphic accelerators. Using analytical modeling and empirical characterization of three real accelerators, we established three neuromorphic bottleneck states (memory, compute, and traffic), and identified which sparsity and parallelization configurations that are likely to be in each bottleneck state. Based on our modeling insights, we synthesized neuromorphic accelerator bounds and bottlenecks into the floorline performance model, analogous to the roofline model for conventional systems, and developed an actionable two-stage optimization framework.


\subsection{Discussion and Future Work}

While we have substantiated the performance boundary trends and optimization methods of the floorline model using extensive characterization and validation on Loihi~2, further in-depth analysis is necessary with other neuromorphic accelerator designs to understand their specific performance bound and bottleneck implications. Currently, such analysis is limited by the research infrastructure and access of nascent neuromorphic accelerators.

Addressing limited research access, further micro-architectural profiling and simulation can inform the construction of an accurate Loihi~2 floorline performance simulator, which can be used as a software proxy for Loihi~2 in network training and compiling research. Such a tool would require not only neurocore-level modeling to inform memory and compute boundaries, but also NoC simulation in order to determine whether workloads have become traffic-bound.

Finally, while the present study has addressed single-chip systems, an important future direction for bottleneck analysis is the extension to multi-chip neuromorphic accelerators, which can include tens to thousands of chips in one system~\cite{halapoint2024, Mayr2019spinnaker2}. Inter-chip communication latency has the potential to dominate the performance over all intra-chip operations, but as inter-chip traffic can also be reduced by network activation sparsity, the relative performance impacts may remain comparable.




\newpage

\bibliographystyle{IEEEtranS}
\bibliography{ref}

\begin{thebibliography}{10}
\providecommand{\url}[1]{#1}
\csname url@samestyle\endcsname
\providecommand{\newblock}{\relax}
\providecommand{\bibinfo}[2]{#2}
\providecommand{\BIBentrySTDinterwordspacing}{\spaceskip=0pt\relax}
\providecommand{\BIBentryALTinterwordstretchfactor}{4}
\providecommand{\BIBentryALTinterwordspacing}{\spaceskip=\fontdimen2\font plus
\BIBentryALTinterwordstretchfactor\fontdimen3\font minus \fontdimen4\font\relax}
\providecommand{\BIBforeignlanguage}[2]{{%
\expandafter\ifx\csname l@#1\endcsname\relax
\typeout{** WARNING: IEEEtranS.bst: No hyphenation pattern has been}%
\typeout{** loaded for the language `#1'. Using the pattern for}%
\typeout{** the default language instead.}%
\else
\language=\csname l@#1\endcsname
\fi
#2}}
\providecommand{\BIBdecl}{\relax}
\BIBdecl

\bibitem{abreu2025neuromorphicprinciplesefficientlarge}
\BIBentryALTinterwordspacing
S.~Abreu, S.~B. Shrestha, R.-J. Zhu, and J.~Eshraghian, ``Neuromorphic principles for efficient large language models on intel loihi 2,'' 2025. [Online]. Available: \url{https://arxiv.org/abs/2503.18002}
\BIBentrySTDinterwordspacing

\bibitem{akopyan15truenorth}
F.~Akopyan, J.~Sawada, A.~Cassidy, R.~Alvarez-Icaza, J.~Arthur, P.~Merolla, N.~Imam, Y.~Nakamura, P.~Datta, G.-J. Nam, B.~Taba, M.~Beakes, B.~Brezzo, J.~B. Kuang, R.~Manohar, W.~P. Risk, B.~Jackson, and D.~S. Modha, ``Truenorth: Design and tool flow of a 65 mw 1 million neuron programmable neurosynaptic chip,'' \emph{IEEE Transactions on Computer-Aided Design of Integrated Circuits and Systems}, vol.~34, no.~10, pp. 1537--1557, 2015.

\bibitem{balaji20spinemap}
A.~Balaji, A.~Das, Y.~Wu, K.~Huynh, F.~G. Dell’Anna, G.~Indiveri, J.~L. Krichmar, N.~D. Dutt, S.~Schaafsma, and F.~Catthoor, ``Mapping spiking neural networks to neuromorphic hardware,'' \emph{IEEE Transactions on Very Large Scale Integration (VLSI) Systems}, vol.~28, no.~1, pp. 76--86, 2020.

\bibitem{basu22snnhwreview}
A.~Basu, L.~Deng, C.~Frenkel, and X.~Zhang, ``Spiking neural network integrated circuits: A review of trends and future directions,'' in \emph{2022 IEEE Custom Integrated Circuits Conference (CICC)}, 2022, pp. 1--8.

\bibitem{bojarski2016endtoend}
\BIBentryALTinterwordspacing
M.~Bojarski, D.~D. Testa, D.~Dworakowski, B.~Firner, B.~Flepp, P.~Goyal, L.~D. Jackel, M.~Monfort, U.~Muller, J.~Zhang, X.~Zhang, J.~Zhao, and K.~Zieba, ``End to end learning for self-driving cars,'' 2016. [Online]. Available: \url{https://arxiv.org/abs/1604.07316}
\BIBentrySTDinterwordspacing

\bibitem{boyle2023sanafe}
\BIBentryALTinterwordspacing
J.~Boyle, M.~Plagge, S.~G. Cardwell, F.~S. Chance, and A.~Gerstlauer, ``Performance and energy simulation of spiking neuromorphic architectures for fast exploration,'' in \emph{Proceedings of the 2023 International Conference on Neuromorphic Systems}, ser. ICONS '23.\hskip 1em plus 0.5em minus 0.4em\relax New York, NY, USA: Association for Computing Machinery, 2023. [Online]. Available: \url{https://doi.org/10.1145/3589737.3605970}
\BIBentrySTDinterwordspacing

\bibitem{brainchipakd1000}
Brainchip, ``Akida akd1000,'' \url{https://brainchip.com/akida-neural-processor-soc/}, 2025.

\bibitem{akidanet}
------, ``Akidanet/imagenet inference,'' \url{https://doc.brainchipinc.com/examples/general/plot_1_akidanet_imagenet.html#sphx-glr-examples-general-plot-1-akidanet-imagenet-py}, 2025.

\bibitem{metatfdocs}
------, ``Metatf documentation,'' \url{https://doc.brainchipinc.com/index.html}, 2025.

\bibitem{chakraborty2024sparse}
\BIBentryALTinterwordspacing
B.~Chakraborty, B.~Kang, H.~Kumar, and S.~Mukhopadhyay, ``Sparse spiking neural network: Exploiting heterogeneity in timescales for pruning recurrent {SNN},'' in \emph{The Twelfth International Conference on Learning Representations}, 2024. [Online]. Available: \url{https://openreview.net/forum?id=0jsfesDZDq}
\BIBentrySTDinterwordspacing

\bibitem{dampfhoffer23aresnn}
M.~Dampfhoffer, T.~Mesquida, A.~Valentian, and L.~Anghel, ``Are snns really more energy-efficient than anns? an in-depth hardware-aware study,'' \emph{IEEE Transactions on Emerging Topics in Computational Intelligence}, vol.~7, no.~3, pp. 731--741, 2023.

\bibitem{davies18loihi}
M.~Davies, N.~Srinivasa, T.-H. Lin, G.~Chinya, Y.~Cao, S.~H. Choday, G.~Dimou, P.~Joshi, N.~Imam, S.~Jain, Y.~Liao, C.-K. Lin, A.~Lines, R.~Liu, D.~Mathaikutty, S.~McCoy, A.~Paul, J.~Tse, G.~Venkataramanan, Y.-H. Weng, A.~Wild, Y.~Yang, and H.~Wang, ``Loihi: A neuromorphic manycore processor with on-chip learning,'' \emph{IEEE Micro}, vol.~38, no.~1, pp. 82--99, 2018.

\bibitem{imagenet}
J.~Deng, W.~Dong, R.~Socher, L.-J. Li, K.~Li, and L.~Fei-Fei, ``Imagenet: A large-scale hierarchical image database,'' in \emph{2009 IEEE Conference on Computer Vision and Pattern Recognition}, 2009, pp. 248--255.

\bibitem{Furber_2016}
\BIBentryALTinterwordspacing
S.~Furber, ``Large-scale neuromorphic computing systems,'' \emph{Journal of Neural Engineering}, vol.~13, no.~5, p. 051001, aug 2016. [Online]. Available: \url{https://dx.doi.org/10.1088/1741-2560/13/5/051001}
\BIBentrySTDinterwordspacing

\bibitem{furber14spinnaker}
S.~B. Furber, F.~Galluppi, S.~Temple, and L.~A. Plana, ``The spinnaker project,'' \emph{Proceedings of the IEEE}, vol. 102, no.~5, pp. 652--665, 2014.

\bibitem{gomez2023first}
W.~G. Gomez, A.~Pignata, R.~Pignari, V.~Fra, E.~Macii, and G.~Urgese, ``First steps towards micro-benchmarking the lava-loihi neuromorphic ecosystem,'' in \emph{2023 IEEE 16th International Symposium on Embedded Multicore/Many-core Systems-on-Chip (MCSoC)}, 2023, pp. 462--469.

\bibitem{howard2017mobilenetsefficientconvolutionalneural}
\BIBentryALTinterwordspacing
A.~G. Howard, M.~Zhu, B.~Chen, D.~Kalenichenko, W.~Wang, T.~Weyand, M.~Andreetto, and H.~Adam, ``Mobilenets: Efficient convolutional neural networks for mobile vision applications,'' 2017. [Online]. Available: \url{https://arxiv.org/abs/1704.04861}
\BIBentrySTDinterwordspacing

\bibitem{imagenette}
\BIBentryALTinterwordspacing
J.~Howard, ``imagenette.'' [Online]. Available: \url{https://github.com/fastai/imagenette/}
\BIBentrySTDinterwordspacing

\bibitem{indiveri00dvs}
\BIBentryALTinterwordspacing
G.~Indiveri and R.~Douglas, ``Neuromorphic vision sensors,'' \emph{Science}, vol. 288, no. 5469, pp. 1189--1190, 2000. [Online]. Available: \url{https://www.science.org/doi/abs/10.1126/science.288.5469.1189}
\BIBentrySTDinterwordspacing

\bibitem{loihi2techbrief}
Intel, ``Taking neuromorphic computing to the next level with loihi 2,'' \url{https://www.intel.com/content/www/us/en/research/neuromorphic-computing-loihi-2-technology-brief.html}, 2021.

\bibitem{halapoint2024}
------, ``Intel builds world’s largest neuromorphic system to enable more sustainable ai,'' \url{https://www.intel.com/content/www/us/en/newsroom/news/intel-builds-worlds-largest-neuromorphic-system.html}, 2024.

\bibitem{lava}
------, ``Lava software framework,'' \url{https://lava-nc.org/}, 2025.

\bibitem{jin23mapping}
\BIBentryALTinterwordspacing
O.~Jin, Q.~Xing, Y.~Li, S.~Deng, S.~He, and G.~Pan, ``Mapping very large scale spiking neuron network to neuromorphic hardware,'' in \emph{Proceedings of the 28th ACM International Conference on Architectural Support for Programming Languages and Operating Systems, Volume 3}, ser. ASPLOS 2023.\hskip 1em plus 0.5em minus 0.4em\relax New York, NY, USA: Association for Computing Machinery, 2023, p. 419–432. [Online]. Available: \url{https://doi.org/10.1145/3582016.3582038}
\BIBentrySTDinterwordspacing

\bibitem{khacef18confronting}
L.~Khacef, N.~Abderrahmane, and B.~Miramond, ``Confronting machine-learning with neuroscience for neuromorphic architectures design,'' in \emph{2018 International Joint Conference on Neural Networks (IJCNN)}, 2018, pp. 1--8.

\bibitem{lazowska1984quantitative}
E.~D. Lazowska, J.~Zahorjan, G.~S. Graham, and K.~C. Sevcik, \emph{Quantitative System Performance: Computer System Analysis Using Queueing Network Models}.\hskip 1em plus 0.5em minus 0.4em\relax Englewood Cliffs, NJ: Prentice‑Hall, Englewood Cliffs, NJ, 1984, originally presented at the CMG International Conference, 1984 (conference paper), later expanded into the book.

\bibitem{li20sneap}
\BIBentryALTinterwordspacing
S.~Li, S.~Guo, L.~Zhang, Z.~Kang, S.~Wang, W.~Shi, L.~Wang, and W.~Xu, ``Sneap: A fast and efficient toolchain for mapping large-scale spiking neural network onto noc-based neuromorphic platform,'' in \emph{Proceedings of the 2020 on Great Lakes Symposium on VLSI}, ser. GLSVLSI '20.\hskip 1em plus 0.5em minus 0.4em\relax New York, NY, USA: Association for Computing Machinery, 2020, p. 9–14. [Online]. Available: \url{https://doi.org/10.1145/3386263.3406900}
\BIBentrySTDinterwordspacing

\bibitem{lin18loihimapping}
\BIBentryALTinterwordspacing
C.-K. Lin, A.~Wild, G.~N. Chinya, T.-H. Lin, M.~Davies, and H.~Wang, ``Mapping spiking neural networks onto a manycore neuromorphic architecture,'' in \emph{Proceedings of the 39th ACM SIGPLAN Conference on Programming Language Design and Implementation}, ser. PLDI 2018.\hskip 1em plus 0.5em minus 0.4em\relax New York, NY, USA: Association for Computing Machinery, 2018, p. 78–89. [Online]. Available: \url{https://doi.org/10.1145/3192366.3192371}
\BIBentrySTDinterwordspacing

\bibitem{man24speck}
Y.~Man, O.~Richter, G.~Zhao, N.~Qiao, Y.~Xing, W.~Dingheng, T.~Hu, W.~Fang, T.~Demirci, M.~Marchi, T.~Yan, C.~Nielsen, S.~Sheik, C.~Wu, Y.~Tian, B.~Xu, and G.~Li, ``Spike-based dynamic computing with asynchronous sensing-computing neuromorphic chip,'' \emph{Nature Communications}, vol.~15, 05 2024.

\bibitem{Mayr2019spinnaker2}
C.~Mayr, S.~Hoeppner, and S.~Furber, ``Spinnaker 2: A 10 million core processor system for brain simulation and machine learning,'' 2019.

\bibitem{meyer2024diagonalstructuredstatespace}
\BIBentryALTinterwordspacing
S.~M. Meyer, P.~Weidel, P.~Plank, L.~Campos-Macias, S.~B. Shrestha, P.~Stratmann, and M.~Richter, ``A diagonal structured state space model on loihi 2 for efficient streaming sequence processing,'' 2024. [Online]. Available: \url{https://arxiv.org/abs/2409.15022}
\BIBentrySTDinterwordspacing

\bibitem{moreira20neuronflow}
O.~Moreira, A.~Yousefzadeh, F.~Chersi, G.~Cinserin, R.-J. Zwartenkot, A.~Kapoor, P.~Qiao, P.~Kievits, M.~Khoei, L.~Rouillard, A.~Ferouge, J.~Tapson, and A.~Visweswara, ``Neuronflow: a neuromorphic processor architecture for live ai applications,'' in \emph{2020 Design, Automation \& Test in Europe Conference \& Exhibition (DATE)}, 2020, pp. 840--845.

\bibitem{na2022autosnnenergyefficientspikingneural}
\BIBentryALTinterwordspacing
B.~Na, J.~Mok, S.~Park, D.~Lee, H.~Choe, and S.~Yoon, ``Autosnn: Towards energy-efficient spiking neural networks,'' 2022. [Online]. Available: \url{https://arxiv.org/abs/2201.12738}
\BIBentrySTDinterwordspacing

\bibitem{nazeer24languagesp2}
K.~K. Nazeer, M.~Schöne, R.~Mukherji, B.~Vogginger, C.~Mayr, D.~Kappel, and A.~Subramoney, ``Language modeling on a spinnaker2 neuromorphic chip,'' in \emph{2024 IEEE 6th International Conference on AI Circuits and Systems (AICAS)}, 2024, pp. 492--496.

\bibitem{o'connor2017sigma}
\BIBentryALTinterwordspacing
P.~O'Connor and M.~Welling, ``Sigma delta quantized networks,'' in \emph{International Conference on Learning Representations}, 2017. [Online]. Available: \url{https://openreview.net/forum?id=HkNRsU5ge}
\BIBentrySTDinterwordspacing

\bibitem{orchard2015converting}
G.~Orchard, A.~Jayawant, G.~K. Cohen, and N.~Thakor, ``Converting static image datasets to spiking neuromorphic datasets using saccades,'' \emph{Frontiers in neuroscience}, vol.~9, p. 437, 2015.

\bibitem{Ostrau2022}
\BIBentryALTinterwordspacing
C.~Ostrau, C.~Klarhorst, M.~Thies, and U.~Rückert, ``Benchmarking neuromorphic hardware and its energy expenditure,'' \emph{Frontiers in Neuroscience}, vol.~16, 2022. [Online]. Available: \url{https://www.frontiersin.org/articles/10.3389/fnins.2022.873935}
\BIBentrySTDinterwordspacing

\bibitem{pei2019tianjic}
J.~Pei, L.~Deng, S.~Song, M.~Zhao, Y.~Zhang, S.~Wu, G.~Wang, Z.~Zou, Z.~Wu, W.~He \emph{et~al.}, ``Towards artificial general intelligence with hybrid tianjic chip architecture,'' \emph{Nature}, vol. 572, no. 7767, pp. 106--111, 2019.

\bibitem{pierro2025acceleratinglinearrecurrentneural}
\BIBentryALTinterwordspacing
A.~Pierro, S.~Abreu, J.~Timcheck, P.~Stratmann, A.~Wild, and S.~B. Shrestha, ``Accelerating linear recurrent neural networks for the edge with unstructured sparsity,'' 2025. [Online]. Available: \url{https://arxiv.org/abs/2502.01330}
\BIBentrySTDinterwordspacing

\bibitem{richter2024speck}
\BIBentryALTinterwordspacing
O.~Richter, Y.~Xing, M.~D. Marchi, C.~Nielsen, M.~Katsimpris, R.~Cattaneo, Y.~Ren, Y.~Hu, Q.~Liu, S.~Sheik, T.~Demirci, and N.~Qiao, ``Speck: A smart event-based vision sensor with a low latency 327k neuron convolutional neuronal network processing pipeline,'' 2024. [Online]. Available: \url{https://arxiv.org/abs/2304.06793}
\BIBentrySTDinterwordspacing

\bibitem{roy2019towards}
K.~Roy, A.~Jaiswal, and P.~Panda, ``Towards spike-based machine intelligence with neuromorphic computing,'' \emph{Nature}, vol. 575, no. 7784, pp. 607--617, 2019.

\bibitem{schuman2017surveyneuromorphiccomputingneural}
\BIBentryALTinterwordspacing
C.~D. Schuman, T.~E. Potok, R.~M. Patton, J.~D. Birdwell, M.~E. Dean, G.~S. Rose, and J.~S. Plank, ``A survey of neuromorphic computing and neural networks in hardware,'' 2017. [Online]. Available: \url{https://arxiv.org/abs/1705.06963}
\BIBentrySTDinterwordspacing

\bibitem{severa25benchmarkingpartitioning}
\BIBentryALTinterwordspacing
W.~Severa, F.~Wang, Y.~Ho, F.~Rothganger, A.~Daram, and E.~Gonzalez, ``Benchmarking spiking network partitioning methods on loihi 2,'' in \emph{Proceedings of the Great Lakes Symposium on VLSI 2025}, ser. GLSVLSI '25.\hskip 1em plus 0.5em minus 0.4em\relax New York, NY, USA: Association for Computing Machinery, 2025, p. 898–904. [Online]. Available: \url{https://doi.org/10.1145/3716368.3735294}
\BIBentrySTDinterwordspacing

\bibitem{shen23esl-snn}
\BIBentryALTinterwordspacing
J.~Shen, Q.~Xu, J.~K. Liu, Y.~Wang, G.~Pan, and H.~Tang, ``Esl-snns: an evolutionary structure learning strategy for spiking neural networks,'' in \emph{Proceedings of the Thirty-Seventh AAAI Conference on Artificial Intelligence and Thirty-Fifth Conference on Innovative Applications of Artificial Intelligence and Thirteenth Symposium on Educational Advances in Artificial Intelligence}, ser. AAAI'23/IAAI'23/EAAI'23.\hskip 1em plus 0.5em minus 0.4em\relax AAAI Press, 2023. [Online]. Available: \url{https://doi.org/10.1609/aaai.v37i1.25079}
\BIBentrySTDinterwordspacing

\bibitem{shi2024towards}
\BIBentryALTinterwordspacing
X.~Shi, J.~Ding, Z.~Hao, and Z.~Yu, ``Towards energy efficient spiking neural networks: An unstructured pruning framework,'' in \emph{The Twelfth International Conference on Learning Representations}, 2024. [Online]. Available: \url{https://openreview.net/forum?id=eoSeaK4QJo}
\BIBentrySTDinterwordspacing

\bibitem{shoesmith2025eventproptrainingefficientneuromorphic}
\BIBentryALTinterwordspacing
T.~Shoesmith, J.~C. Knight, B.~Mészáros, J.~Timcheck, and T.~Nowotny, ``Eventprop training for efficient neuromorphic applications,'' 2025. [Online]. Available: \url{https://arxiv.org/abs/2503.04341}
\BIBentrySTDinterwordspacing

\bibitem{shrestha2024efficient}
S.~B. Shrestha, J.~Timcheck, P.~Frady, L.~Campos-Macias, and M.~Davies, ``Efficient video and audio processing with loihi 2,'' in \emph{ICASSP 2024 - 2024 IEEE International Conference on Acoustics, Speech and Signal Processing (ICASSP)}, 2024, pp. 13\,481--13\,485.

\bibitem{smith2023simplifiedstatespacelayers}
\BIBentryALTinterwordspacing
J.~T.~H. Smith, A.~Warrington, and S.~W. Linderman, ``Simplified state space layers for sequence modeling,'' 2023. [Online]. Available: \url{https://arxiv.org/abs/2208.04933}
\BIBentrySTDinterwordspacing

\bibitem{smith2023simplified}
\BIBentryALTinterwordspacing
J.~T. Smith, A.~Warrington, and S.~Linderman, ``Simplified state space layers for sequence modeling,'' in \emph{The Eleventh International Conference on Learning Representations}, 2023. [Online]. Available: \url{https://openreview.net/forum?id=Ai8Hw3AXqks}
\BIBentrySTDinterwordspacing

\bibitem{song22dfsynthesizer}
\BIBentryALTinterwordspacing
S.~Song, H.~Chong, A.~Balaji, A.~Das, J.~Shackleford, and N.~Kandasamy, ``Dfsynthesizer: Dataflow-based synthesis of spiking neural networks to neuromorphic hardware,'' \emph{ACM Trans. Embed. Comput. Syst.}, vol.~21, no.~3, May 2022. [Online]. Available: \url{https://doi.org/10.1145/3479156}
\BIBentrySTDinterwordspacing

\bibitem{sorbaro20synop_loss}
\BIBentryALTinterwordspacing
M.~Sorbaro, Q.~Liu, M.~Bortone, and S.~Sheik, ``Optimizing the energy consumption of spiking neural networks for neuromorphic applications,'' \emph{Frontiers in Neuroscience}, vol. Volume 14 - 2020, 2020. [Online]. Available: \url{https://www.frontiersin.org/journals/neuroscience/articles/10.3389/fnins.2020.00662}
\BIBentrySTDinterwordspacing

\bibitem{sinabs}
Synsense, ``Sinabs is not a brain simulator,'' \url{https://sinabs.readthedocs.io/}, 2025.

\bibitem{tang2023openboxdigitalneuromorphic}
\BIBentryALTinterwordspacing
G.~Tang, A.~Safa, K.~Shidqi, P.~Detterer, S.~Traferro, M.~Konijnenburg, M.~Sifalakis, G.-J. van Schaik, and A.~Yousefzadeh, ``Open the box of digital neuromorphic processor: Towards effective algorithm-hardware co-design,'' 2023. [Online]. Available: \url{https://arxiv.org/abs/2303.15224}
\BIBentrySTDinterwordspacing

\bibitem{tang2023seneca}
G.~Tang, K.~Vadivel, Y.~Xu, R.~Bilgic, K.~Shidqi, P.~Detterer, S.~Traferro, M.~Konijnenburg, M.~Sifalakis, G.-J. van Schaik \emph{et~al.}, ``Seneca: building a fully digital neuromorphic processor, design trade-offs and challenges,'' \emph{Frontiers in Neuroscience}, vol.~17, p. 1187252, 2023.

\bibitem{Timcheck_2023}
\BIBentryALTinterwordspacing
J.~Timcheck, S.~B. Shrestha, D.~Ben Dayan~Rubin, A.~Kupryjanow, G.~Orchard, L.~Pindor, T.~Shea, and M.~Davies, ``The intel neuromorphic dns challenge,'' \emph{Neuromorphic Computing and Engineering}, vol.~3, no.~3, p. 034005, aug 2023. [Online]. Available: \url{https://dx.doi.org/10.1088/2634-4386/ace737}
\BIBentrySTDinterwordspacing

\bibitem{urgese2016optimizing}
G.~Urgese, F.~Barchi, E.~Macii, and A.~Acquaviva, ``Optimizing network traffic for spiking neural network simulations on densely interconnected many-core neuromorphic platforms,'' \emph{IEEE Transactions on Emerging Topics in Computing}, vol.~6, no.~3, pp. 317--329, 2016.

\bibitem{williams09roofline}
\BIBentryALTinterwordspacing
S.~Williams, A.~Waterman, and D.~Patterson, ``Roofline: an insightful visual performance model for multicore architectures,'' \emph{Commun. ACM}, vol.~52, no.~4, p. 65–76, Apr. 2009. [Online]. Available: \url{https://doi.org/10.1145/1498765.1498785}
\BIBentrySTDinterwordspacing

\bibitem{wu20rlmapping}
\BIBentryALTinterwordspacing
N.~Wu, L.~Deng, G.~Li, and Y.~Xie, ``Core placement optimization for multi-chip many-core neural network systems with reinforcement learning,'' \emph{ACM Trans. Des. Autom. Electron. Syst.}, vol.~26, no.~2, Oct. 2020. [Online]. Available: \url{https://doi.org/10.1145/3418498}
\BIBentrySTDinterwordspacing

\bibitem{xie23spikenc}
L.~Xie, J.~Xue, L.~Wu, F.~Chen, Q.~Tian, Y.~Zhou, R.~Ying, and P.~Liu, ``Spikenc: An accurate and scalable simulator for spiking neural network on multi-core neuromorphic hardware,'' in \emph{2023 IEEE 30th International Conference on High Performance Computing, Data, and Analytics (HiPC)}, 2023, pp. 357--366.

\bibitem{imec2024optimizingseneca}
\BIBentryALTinterwordspacing
Y.~Xu, K.~Shidqi, G.-J. van Schaik, R.~Bilgic, A.~Dobrita, S.~Wang, R.~Meijer, P.~Nembhani, C.~Arjmand, P.~Martinello, A.~Gebregiorgis, S.~Hamdioui, P.~Detterer, S.~Traferro, M.~Konijnenburg, K.~Vadivel, M.~Sifalakis, G.~Tang, and A.~Yousefzadeh, ``Optimizing event-based neural networks on digital neuromorphic architecture: a comprehensive design space exploration,'' \emph{Frontiers in Neuroscience}, vol. Volume 18 - 2024, 2024. [Online]. Available: \url{https://www.frontiersin.org/journals/neuroscience/articles/10.3389/fnins.2024.1335422}
\BIBentrySTDinterwordspacing

\bibitem{yan2022sparsityregularization}
\BIBentryALTinterwordspacing
Y.~Yan, H.~Chu, Y.~Jin, Y.~Huan, Z.~Zou, and L.~Zheng, ``Backpropagation with sparsity regularization for spiking neural network learning,'' \emph{Frontiers in Neuroscience}, vol.~16, 2022. [Online]. Available: \url{https://www.frontiersin.org/journals/neuroscience/articles/10.3389/fnins.2022.760298}
\BIBentrySTDinterwordspacing

\bibitem{yang24tianjiccompiling}
Y.~Yang, Q.~Fan, T.~Yan, J.~Pei, and G.~Li, ``Network group partition and core placement optimization for neuromorphic multi-core and multi-chip systems,'' \emph{IEEE Transactions on Emerging Topics in Computational Intelligence}, vol.~8, no.~6, pp. 3966--3981, 2024.

\bibitem{yik2025neurobench}
J.~{Yik}, K.~{Van den Berghe}, D.~{den Blanken}, Y.~{Bouhadjar}, M.~{Fabre}, P.~{Hueber}, W.~{Ke}, M.~A. {Khoei}, D.~{Kleyko}, N.~{Pacik-Nelson}, A.~{Pierro}, P.~{Stratmann}, P.-S.~V. {Sun}, G.~{Tang}, S.~{Wang}, B.~{Zhou}, S.~H. {Ahmed}, G.~{Vathakkattil Joseph}, B.~{Leto}, A.~{Micheli}, A.~K. {Mishra}, G.~{Lenz}, T.~{Sun}, Z.~{Ahmed}, M.~{Akl}, B.~{Anderson}, A.~G. {Andreou}, C.~{Bartolozzi}, A.~{Basu}, P.~{Bogdan}, S.~{Bohte}, S.~{Buckley}, G.~{Cauwenberghs}, E.~{Chicca}, F.~{Corradi}, G.~{de Croon}, A.~{Danielescu}, A.~{Daram}, M.~{Davies}, Y.~{Demirag}, J.~{Eshraghian}, T.~{Fischer}, J.~{Forest}, V.~{Fra}, S.~{Furber}, P.~M. {Furlong}, W.~{Gilpin}, A.~{Gilra}, H.~A. {Gonzalez}, G.~{Indiveri}, S.~{Joshi}, V.~{Karia}, L.~{Khacef}, J.~C. {Knight}, L.~{Kriener}, R.~{Kubendran}, D.~{Kudithipudi}, S.-C. {Liu}, Y.-H. {Liu}, H.~{Ma}, R.~{Manohar}, J.~M. {Margarit-Taul{\'e}}, C.~{Mayr}, K.~{Michmizos}, D.~R. {Muir}, E.~{Neftci}, T.~{Nowotny}, F.~{Ottati}, A.~{Ozcelikkale}, P.~{Panda}, J.~{Park}, M.~{Payvand},
  C.~{Pehle}, M.~A. {Petrovici}, C.~{Posch}, A.~{Renner}, Y.~{Sandamirskaya}, C.~J.~S. {Schaefer}, A.~{van Schaik}, J.~{Schemmel}, S.~{Schmidgall}, C.~{Schuman}, J.-s. {Seo}, S.~{Sheik}, S.~B. {Shrestha}, M.~{Sifalakis}, A.~{Sironi}, K.~{Stewart}, M.~{Stewart}, T.~C. {Stewart}, J.~{Timcheck}, N.~{T{\"o}men}, G.~{Urgese}, M.~{Verhelst}, C.~M. {Vineyard}, B.~{Vogginger}, A.~{Yousefzadeh}, F.~T. {Zohora}, C.~{Frenkel}, and V.~J. {Reddi}, ``{The neurobench framework for benchmarking neuromorphic computing algorithms and systems},'' \emph{Nature Communications}, vol.~16, no.~1, p. 1545, Feb. 2025.

\bibitem{yu2024dualsparsetrainingframework}
\BIBentryALTinterwordspacing
X.~Yu and C.~Tian, ``Dual sparse training framework: inducing activation map sparsity via transformed $\ell1$ regularization,'' 2024. [Online]. Available: \url{https://arxiv.org/abs/2405.19652}
\BIBentrySTDinterwordspacing

\end{thebibliography}

\end{document}